\documentclass[twocolumn]{aastex62}

\usepackage{natbib}
\usepackage{amsfonts}
\usepackage{amsmath}
\usepackage{graphicx}
\usepackage{xspace}
\usepackage{mathrsfs}
\usepackage{bm}
\usepackage[normalem]{ulem}

\hypersetup{linkcolor=red,citecolor=magenta,filecolor=cyan,urlcolor=blue}

\newcommand\teff{T_{\rm eff}}
\newcommand\tint{T_{\rm int}}
\newcommand\teq{T_{\rm eq}}
\newcommand\tone{T_{\rm 1}}

\newcommand\grad\nabla

\newcommand\grada{\nabla_{\rm ad}}

\newcommand{\xhe}{x_{\rm He}}
\newcommand{\pr}{\rm Pr}
\newcommand{\kmu}{\kappa_\mu}
\newcommand{\kt}{\kappa_T}

\newcommand{\mc}{M_{\rm c}}
\newcommand{\dt}{\Delta T_{\rm phase}}
\newcommand{\rrho}{R_\rho}

\newcommand{\tsun}{\tau_\odot}

\newcommand\sch{SR18\xspace}%{Sch\"ottler}
\newcommand{\juno}{\emph{Juno}\xspace}
\newcommand{\cassini}{\emph{Cassini}\xspace}
\newcommand{\voyager}{\emph{Voyager}\xspace}
\newcommand{\galileo}{\emph{Galileo}\xspace}
\newcommand{\pioneer}{\emph{Pioneer}\xspace}

\accepted{December 14, 2019}
\submitjournal{ApJ}

\begin{document}
\shortauthors{Mankovich and Fortney}
\shorttitle{Helium in Jupiter and Saturn}
\title{Evidence for a Dichotomy in the Interior Structures of Jupiter and Saturn from Helium Phase Separation}

\correspondingauthor{Chris Mankovich}
\email{chkvch@caltech.edu}

\newcommand\ucsc{University of California Santa Cruz, Department of Astronomy and Astrophysics, Santa Cruz, CA 95064, USA}
\newcommand\caltech{California Institute of Technology, Division of Geological and Planetary Sciences, Pasadena, CA 91125, USA}

\author{Christopher R. Mankovich}
\affiliation\ucsc\affiliation\caltech

\author{Jonathan J. Fortney}
\affiliation\ucsc

\begin{abstract}
  We examine the comparative thermal evolution of Jupiter and Saturn applying recent theoretical results for helium's immiscibility in fluid metallic hydrogen. The redistribution of helium in their interiors proceeds very differently for the two planets. We confirm that based on Jupiter's atmospheric helium depletion as observed \textit{in situ} by the \galileo entry probe, Jupiter's interior helium has differentiated modestly, and we present models reconciling Jupiter's helium depletion, radius, and heat flow at the solar age. Jupiter's recently revised Bond albedo implies a lower intrinsic flux for the planet, accommodating more luminosity from helium differentiation such that mildly superadiabatic interiors can satisfy all constraints. The same phase diagram applied to the less massive Saturn predicts dramatic helium differentiation to the degree that Saturn inevitably forms a helium-rich shell or core, an outcome previously proposed by Stevenson \& Salpeter and others. The luminosity from Saturn's helium differentiation is sufficient to extend its cooling time to the solar age, even for adiabatic interiors. This model predicts Saturn's atmospheric helium to be depleted to $Y=0.07\pm0.01$, corresponding to a He/H$_2$ mixing ratio $0.036\pm0.006$. We also show that neon differentiation may have contributed to both planets' luminosity in the past. These results demonstrate that Jupiter and Saturn's thermal evolution can be explained self-consistently with a single physical model, and {emphasize that nontrivial helium distributions should be considered in future models for Saturn's internal structure and dynamo.}
\end{abstract}

\keywords{}

\section{Introduction}\label{s.intro}
Understanding the interiors of the gas giants is a critical step toward understanding the universal processes of planet formation and evolution. Jupiter and Saturn hold special significance in this respect because of their accessibility. However, outstanding puzzles concerning their thermal evolution obscure the connection between their present-day configurations and their origins in the young solar system. Evolutionary models treating Jupiter's interior as being well-mixed and nearly adiabatic as a result of efficient convection are broadly successful in explaining the planet's luminosity at the solar age \citep{1975ApJ...199..265G,fortney2011}. However, similar models for Saturn fail to reproduce its observed heat flow \citep{1977Icar...30..111P,1980Icar...42..358G,fortney2011}, and thus some additional luminosity source is required.

Apart from the primary luminosity derived from the thermal energy of the interior \citep{1968ApJ...152..745H}, differentiation has long been appreciated as a potentially significant luminosity source for cool gas giants \citep{1967Natur.215..691S,1973ApJ...186.1097F}.
In particular, the limited solubility of helium in fluid metallic hydrogen eventually leads to the formation of helium-rich droplets that may rain out on a timescale short compared to their convective redistribution \citep{1973ApJ...181L..83S}.
As noted by \cite{ss77b}, the success of homogeneous, adiabatic Jupiter models implies that the planet has begun raining out helium only recently or not at all, whereas the differentiation is probably significant in the cooler Saturn.
The helium depletion subsequently observed in Jupiter's atmosphere \citep{1998JGR...10322815V} suggests that the planet has indeed begun differentiating helium in the recent past.

The notion that helium rain can explain Saturn's luminosity is supported by evolutionary calculations including helium immiscibility for plausible, if tentative, phase diagrams \citep{1999P&SS...47.1175H,2003Icar..164..228F} (see also the review by \citealt{2016arXiv160906324F}). \cite{2013NatGe...6..347L} imagined an important alternative scenario wherein significant departures from adiabaticity due to double-diffusive convection in Saturn's deep interior can also explain that planet's luminosity without recourse to helium immiscibility. However, given the direct evidence for helium differentiation in Jupiter, it appears difficult to avoid helium differentiation in the presumably\footnote{
% This presumption would break down if Saturn's interior were superadiabatic enough that its metallic layers were warmer than those of Jupiter in spite of Saturn's colder surface.
For reference, although the baseline layered-convection Saturn model of Leconte \& Chabrier shows a radically different cooling history from their adiabatic case, the two possess very similar deep temperatures at the solar age (their Figure 6), i.e., colder than Jupiter's.} colder interior of Saturn.

Work in recent years has applied many of these ideas to the evolution of Jupiter \citep{2015MNRAS.447.3422N,2016ApJ...832..113M} and Saturn \citep{2016Icar..267..323P}. The main goal of the present work is to simultaneously study the evolution of Jupiter and Saturn under a single model for hydrogen-helium immiscibility to judge whether a consistent picture exists for their evolution. The motivation for doing this now is twofold.
First, Jupiter's Bond albedo has recently been dramatically revised following analysis of multi-instrument \cassini data, indicating less absorbed solar flux and more internal flux emanating from Jupiter \citep{2018NatCo...9.3709L} than long thought based on a combination of \voyager and \pioneer data \citep{1981JGR....86.8705H}. This updated surface condition implies a greater flux contribution from the interior, potentially attributable to helium rain.
The second motivation is the recent phase diagram of \cite{2018PhRvL.120k5703S}, which builds on prior work \citep[e.g.,][]{2011PhRvB..84w5109L,2013PhRvB..87q4105M} by both including nonideal entropy effects and covering the full range of possible helium fractions.
As will be discussed below, this knowledge of the phase diagram over all mixtures (from helium-poor to helium-rich) is of critical importance for modeling the helium distribution within Saturn. We aim to assess to what degree this proposed phase diagram is viable in the context of Jupiter's atmospheric helium content and Jupiter and Saturn's radius and heat flow at the present epoch.

\section{Hydrogen-helium mixtures}\label{s.phase}
The phase diagram that describes the solubility of helium in fluid metallic hydrogen is uncertain in Jovian interiors, a regime that is difficult to access experimentally. The phase diagram in this regime has been increasingly mapped out by \textit{ab initio} methods, in particular density functional theory--molecular dynamics (``DFT-MD'') simulations used to predict the thermodynamic conditions for helium phase separation. Advances along these lines have been made in recent years \citep{2009PhRvL.102k5701L,2011PhRvB..84w5109L,2009PNAS..106.1324M,2013PhRvB..87q4105M,2018PhRvL.120k5703S}, the results remaining substantially uncertain because they are sensitive to the assumed electron density functional and the accuracy with which the entropy of mixing between hydrogen and helium is treated.

As described by \cite{2013PhRvB..87q4105M}, the nonideal contributions to this entropy of mixing is crucial for satisfying experimental results for molecular hydrogen, and strongly affects predictions for solubility in metallic hydrogen. Because the most recent studies of helium phase separation in the evolution of Jupiter \citep{2015MNRAS.447.3422N,2016ApJ...832..113M} and Saturn \citep{2016Icar..267..323P} made use of results assuming an ideal mixing entropy \citep{2011PhRvB..84w5109L}, the more accurate phase diagram of \cite{2018PhRvL.120k5703S} warrants a reappraisal of this type of model.

\subsection{Modeling the interior helium distributions}
The central assumption of the present work is that the helium distributions are dictated by their instantaneous thermodynamic equilibrium profiles. This amounts to the assumption that metallic regions cooled to the point of becoming supersaturated lose their excess helium instantaneously, reducing the ambient abundance to the saturation value while sinking the He-rich phase ($Y\gtrsim0.9$; see Figure~\ref{fig.sat_phase}) deeper into the planet where it redissolves into the background if possible. These ideas have already been described in the literature \citep{ss77b,2003Icar..164..228F,2015MNRAS.447.3422N,2016Icar..267..323P}, and the algorithm we use in practice is described in detail in \cite{2016ApJ...832..113M}. The major differences from that work are the application of the \citet[][hereafter ``\sch'']{2018PhRvL.120k5703S} phase diagram and extension of these models to the case of Saturn.

\subsection{The phase boundary is a surface $T(P, Y)$}\label{s.mixtures}
The earlier DFT-MD simulations of \cite{2013PhRvB..87q4105M} did derive the full nonideal entropy of hydrogen-helium mixtures using thermodynamic integrations, and this phase diagram has previously been applied to detailed models for the \emph{static structure} of Jupiter \citep{2016ApJ...820...80H} and Saturn \citep{2019Sci...364.2965I}. Its major limitation  is its lack of coverage in mixture space, the published phase diagram being restricted to a single helium number fraction $\xhe=8\%$ representing the protosolar helium abundance (ostensibly the mean abundance of the gaseous jovian envelopes). While this is appropriate for predicting whether and where phase separation will set in for a planet initially well mixed at the protosolar helium abundance, it does not generally yield enough information to calculate the resulting helium distributions in any detail.

\begin{figure}[t]
    \begin{center}
        \includegraphics[width=1.0\columnwidth]{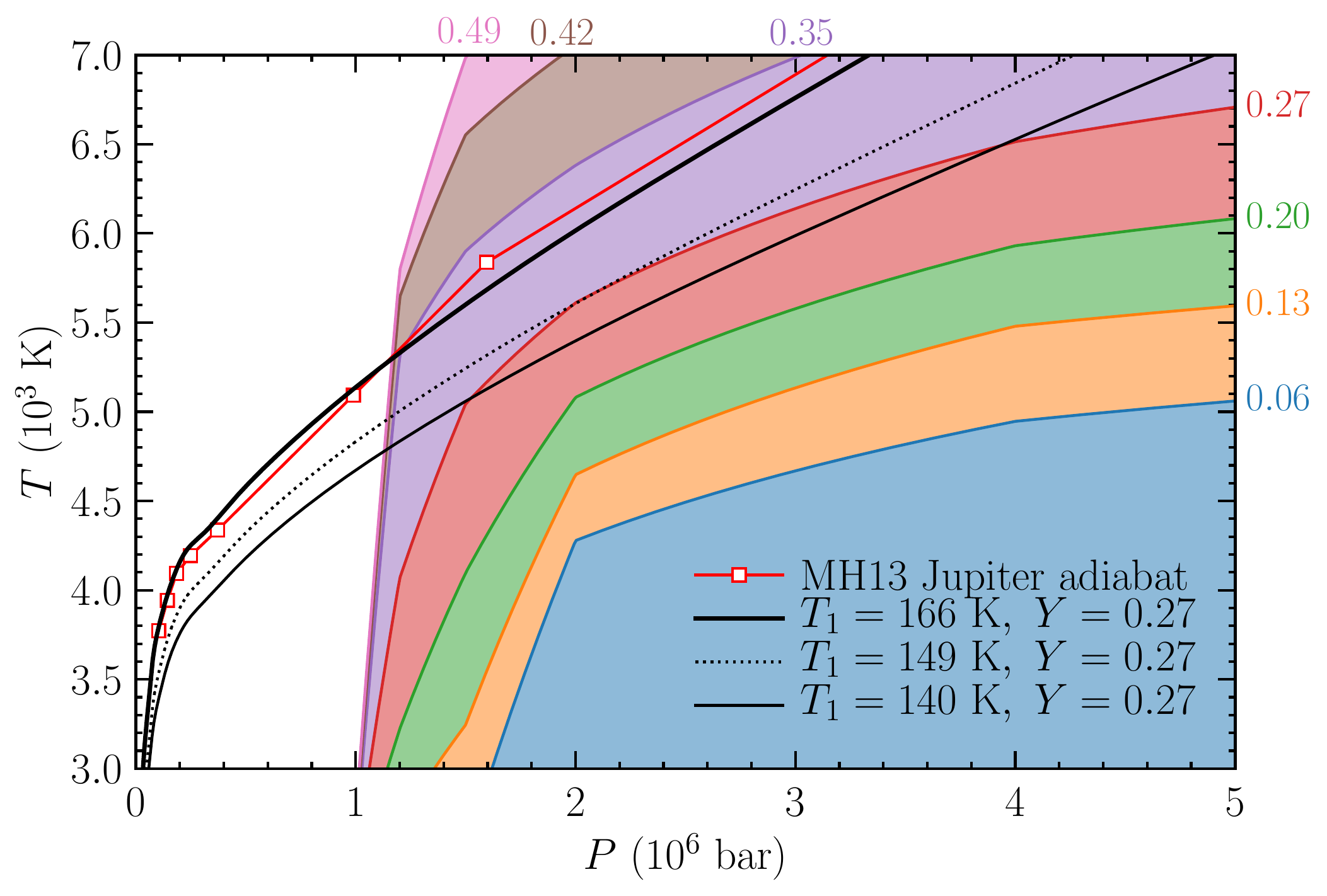} % {../figures/sch_pt}
        \caption{
        Phase curves from the \textit{ab initio} hydrogen-helium phase diagram of \cite{2018PhRvL.120k5703S} compared with protosolar-abundance adiabats (solid black curves) satisfying one-bar temperatures $\tone=166$ and $140$ K corresponding to Jupiter and Saturn, respectively.
        Helium phase separation occurs in regions of the planet that cool beneath the pressure-temperature phase curve for the relevant local helium mass fraction $Y$, color-coded and labeled to the right or top of each phase curve. The diagram focuses on the helium-poor, low-pressure part of the phase diagram relevant for setting the observable helium content of a well-mixed molecular envelope in Jupiter or Saturn. A single intermediate-temperature adiabat ($\tone=149\ {\rm K}$; dotted black curve) represents the onset of helium immiscibility. That Jupiter is $\gtrsim500\ {\rm K}$ warmer than the $\approx5500\ {\rm K}$ onset temperature at $P=2\ {\rm Mbar}$ implies that phase separation would not take place in Jupiter, at odds with its atmospheric helium depletion.
        }
        \label{fig.phase_pt}
    \end{center}
\end{figure}

The basic reason for this is that if a region becomes supercooled and loses its excess helium to greater depths via dense droplets, the local helium abundance decreases to the value satisfying exact saturation. Solving for this saturation abundance requires knowledge of the phase curves corresponding to \emph{lower} abundances than the initial value. Take for example the homogeneous protosolar-abundance adiabats indicated in Figure~\ref{fig.phase_pt}, where they are compared to phase curves obtained by B-spline fits to the \sch data.
The $\tone=149\ {\rm K}$ adiabat indicated by the dotted curve osculates the $Y=0.27$ phase curve at $P\approx2\ {\rm Mbar}$ and $T\approx5500\ {\rm K}$, representing the moment that helium immiscibility sets in within the planet. From this point the initially well-mixed adiabat is supercooled in the neighborhood of $P\approx2\ {\rm Mbar}$ and this region will tend to lose its excess helium to lower depths via droplets. Exterior to this region, the molecular envelope is kept well-mixed by convection and thus the rainout process at $P\approx2\ {\rm Mbar}$ drains helium from exterior regions uniformly. The outer envelope abundance is thus given by the condition of saturation at $P\approx2\ {\rm Mbar}$, i.e., the value of $Y$ labeling the unique phase curve that osculates the planetary $P\textrm{--}T$ profile there.

Meanwhile at depth, droplets descending from above encounter increasingly warm surroundings, eventually redissolving into the medium (unless they reach the core or center of the planet first, a possibility discussed below). Here the helium abundance in the mixture increases. The question of whether this layer is itself now supersaturated requires knowledge of the phase curve at this \emph{greater} local abundance. It becomes evident that, under the assumption that all excess helium is rained out to lower depths and redissolved at its first opportunity, solving for the equilibrium helium distribution throughout the interior is an iterative process that requires knowledge of the phase diagram at a potentially broad range of helium fractions. In other words, the deep abundances are fundamentally not determined locally, and so the extent of the helium rain region cannot be determined from any single phase curve.

Jupiter's molecular envelope helium depletion relative to the protosolar abundance is modest, and the large mass ratio between Jupiter's metallic and molecular regions guarantees only a subtle helium enrichment of the deep interior. Saturn, on the other hand, has lower internal temperatures and so tends to suffer more dramatic differentiation of helium. We find that applying realistic phase diagrams to Saturn's present-day interior produces helium distributions that differ qualitatively from those obtained for Jupiter. This is at odds with the recent Saturn interior models of \cite{2019Sci...364.2965I} and \cite{2019ApJ...879...78M}, which exhibit helium distributions similar to those expected within Jupiter: a uniformly depleted molecular envelope that gradually transitions to a uniform, moderately enriched metallic envelope deeper down. As we will show below, we find that Saturn is cold enough that such an enriched inner envelope would itself be unstable to further phase separation when the phase diagram is queried at the correct (tentative, enriched) abundance as opposed to the initial protosolar abundance.

\begin{figure*}[t]
    \begin{center}
        \includegraphics[height=0.85\textheight]{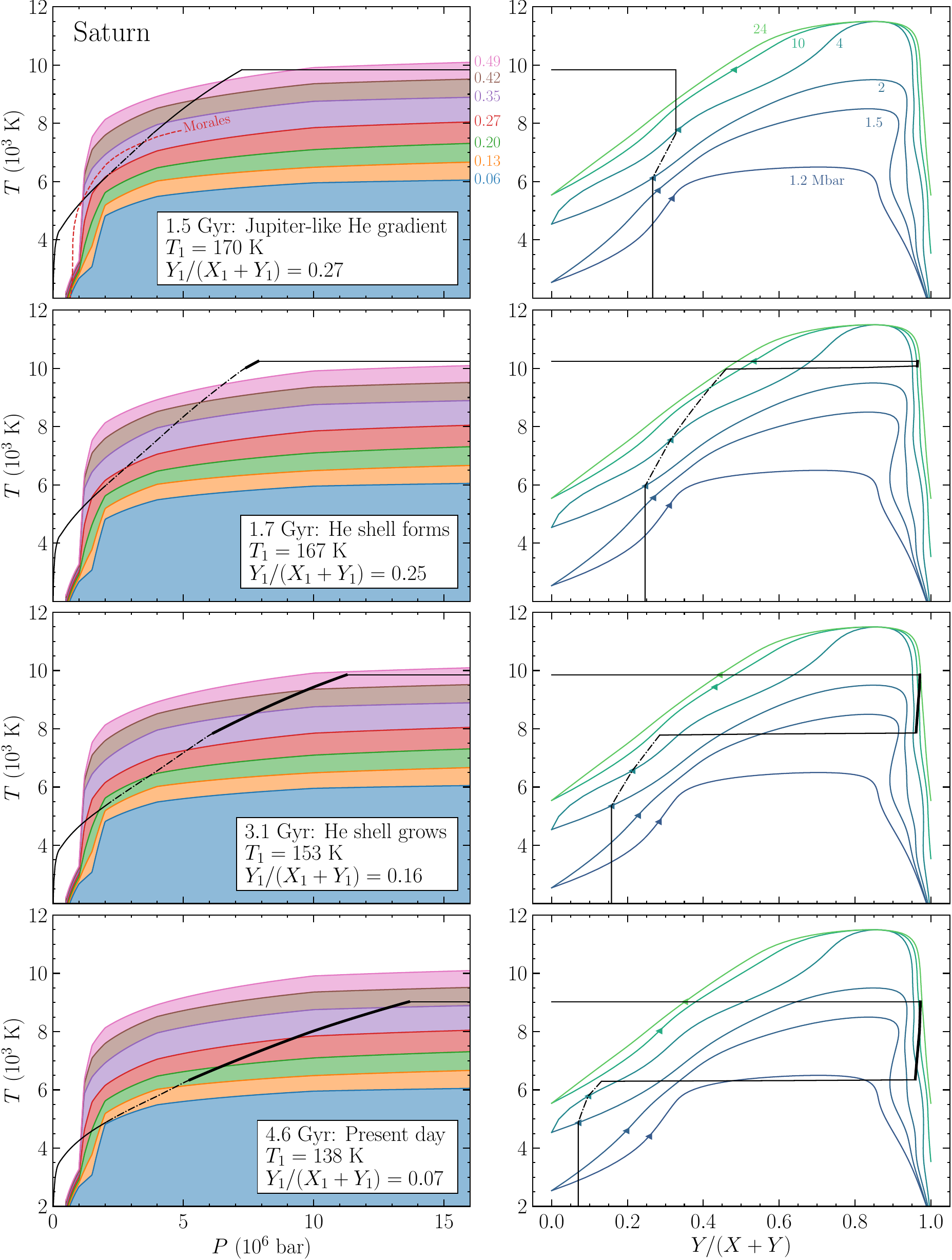} % {../figures/sat_phase-crop}
        \caption{
        A typical sequence in evolutionary time (top to bottom) of Saturn interior profiles (black curves) superimposed with the \sch H-He phase curves (colorful curves){, shifted by $\dt=540\ {\rm K}$ as required to explain Jupiter's helium depletion; see Sections~\ref{s.dtphase} and Figure~\ref{fig.jup_dt_y1} below}. \emph{Left panel:} $P\textrm{--}T$ space. The dot-dashed portion of the planet profile indicates the continuous helium gradient region, and the thick portion indicates the helium-rich shell, after one exists. The \cite{2013PhRvB..87q4105M} phase curve for $Y=0.27$ is included in the topmost panel (dashed curve) for comparison to these shifted \sch phase curves.
        % The thin dotted curve indicates the phase curve corresponding to the instantaneous outer envelope helium abundance. % cut for space
        \emph{Right panel:} the same Saturn profiles in $Y\textrm{--}T$ space, with phase curves corresponding to $P=1.2$, 1.5, 2, 4, 10, and 24 Mbar from bottom/blue to top/green. The triangle on each of these phase curves indicates the maximum $Y$ in the helium-poor phase given the planet's current temperature at that pressure level. These values move to lower $Y$ as the planet cools, driving the depletion in the outer envelope. At 1.7 Gyr the gradient region extends all the way down to the heavy-element ($Y=0$) core, implying that helium-rich material falling from above no longer finds a warm homogeneous inner envelope in which to redissolve. From this time onward material in the helium-rich phase collects outside the core, establishing a dense shell.
        }
        \label{fig.sat_phase}
    \end{center}
\end{figure*}

Speaking in terms of an evolutionary path, we find that after immiscibility sets in, Saturn rapidly cools through a sequence of qualitatively Jupiter-like helium distributions until its helium gradient reaches the planet's dense core (or center, in the absence of such a core). After this point Saturn accumulates a shell (or core, if no core of denser material exists) of helium-rich material, an outcome of hydrogen-helium immiscibility discussed by \cite{1973ApJ...181L..83S} and \citet[][their Figure~4c]{ss77b} and modeled explicitly by \cite{2003Icar..164..228F} and \cite{2016Icar..267..323P}. Figure~\ref{fig.sat_phase} demonstrates this evolution path for our baseline Saturn model with helium rain.

\begin{figure*}[t]
    \begin{center}
        \includegraphics[width=0.55\textwidth]{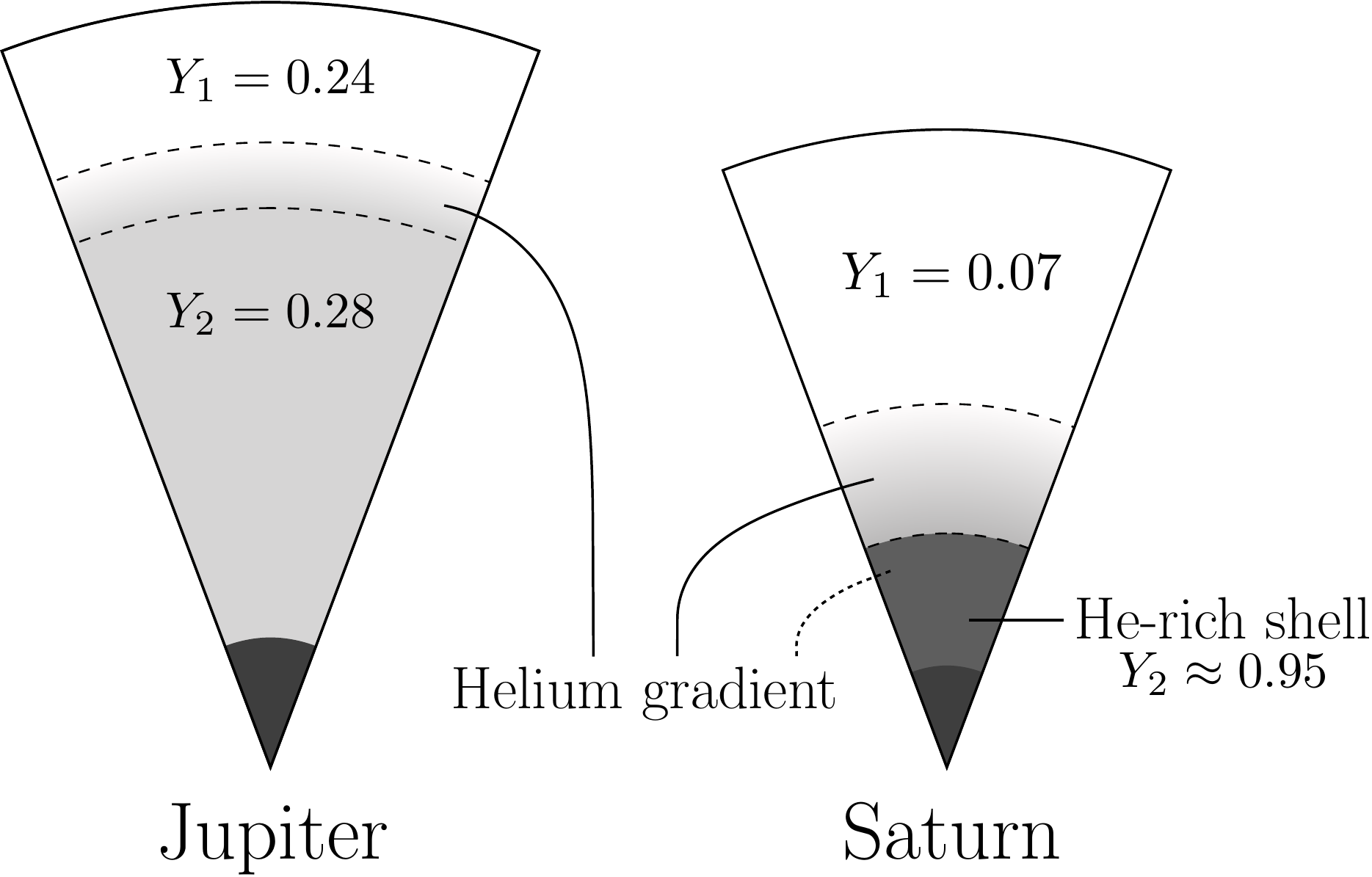} % {../figures/onion-crop}
        \caption{
        A schematic description of the present-day structures found for Jupiter and Saturn by applying the methods described in Section~\ref{s.phase} and \ref{s.models}. Diagrams are to scale by radius, these specific structures corresponding to the most likely individual models from the favored samples described in Section~\ref{s.results}. The dashed boundaries enclose the continuous helium gradient region within each planet. The outer boundaries at $\approx80\%$ of Jupiter's radius and $\approx55\%$ of Saturn's radius correspond to $P\approx2\ {\rm Mbar}$ where \sch predicts the onset of hydrogen-helium immiscibility. The inner boundary at $\approx35\%$ of Saturn's radius represents the transition to a shell of helium-rich material, discussed in Section~\ref{s.phase}. This shell itself possesses a weak helium gradient as can be seen from the right-hand panels of Figure~\ref{fig.sat_phase}. Helium mass fractions quoted here are relative to hydrogen and helium.
        }
        \label{fig.pie_slices}
    \end{center}
\end{figure*}

To help guide the discussion that follows, Figure~\ref{fig.pie_slices} show schematic diagrams representing the typical present-day internal structures that we obtain for Jupiter and Saturn.

\subsection{Overall temperature of the phase diagram}\label{s.dtphase}
{That the $\tone=166\ {\rm K}$ protosolar-abundance adiabat in Figure~\ref{fig.phase_pt} lies well above the $Y=0.27$ phase curve from \sch would suggest that no phase separation occurs in Jupiter, consistent with the findings of \cite{2018PhRvL.120k5703S} who carried out a similar comparison to reference adiabats. Put another way, the $\tone=166\ {\rm K}$ protosolar-abundance adiabat is $\gtrsim500\ {\rm K}$ warmer than the $\approx5500\ {\rm K}$ phase boundary at $Y=0.27$ and $P=2\ {\rm Mbar}$, and thus if phase separation is to explain Jupiter's observed helium depletion, the realistic phase curve must be some $500\ {\rm K}$ warmer than predicted by \sch.}

Assuming that hydrogen-helium immiscibility is the mechanism responsible for Jupiter's atmospheric helium depletion, the \galileo abundance measurement imposes a stringent constraint for discerning among viable, if still uncertain, phase diagrams. It is for this reason that, following \cite{2015MNRAS.447.3422N} and \cite{2016ApJ...832..113M}, we introduce a degree of freedom via an additive temperature offset $\dt$ modulating the overall temperature of the phase curves applied in this work, relative to \sch. This parameter allows us to explore a more general space of phase diagrams, with larger $\dt$ values leading to more pronounced differentiation, smaller values leading to less, and $\dt=0$ recovering the \sch phase curves as published. It also yields a convenient language of expressing our results in terms of belief about the ``true'' phase diagram based on how well our various thermal evolution models fare. Based on the discussion thus far, Jupiter models will require $\dt>0$ to successfully match the \galileo helium abundance.

\section{Gas giant evolution models}\label{s.models}
We create new evolutionary models for Jupiter and Saturn using a code derived from that of \cite{2016ApJ...831...64T} and recently applied to Saturn's static structure in \cite{2019ApJ...871....1M}. The most significant update is the use of the \emph{ab initio} hydrogen-helium equation of state (EOS) of \cite[][``MH13"]{2013ApJ...774..148M}, an advance compared to the semianalytic model of \cite{1995ApJS...99..713S} which predicted warmer metallic interiors for Jupiter and Saturn. MH13 provides data for a single mixture $Y=0.245$. In this work EOS quantities are calculated for arbitrary hydrogen-helium mixtures by combining MH13 with the Saumon et al. table for pure helium under the linear mixing approximation, as described and tabulated by \cite{2016A&A...596A.114M}. Heavier elements are modeled as pure water ice using the Rostock water EOS of \cite{2009PhRvB..79e4107F}, also incorporated under the linear mixing approximation. The models in this work are initialized hot, with uniform envelopes containing a helium mass fraction $Y=0.270$ corresponding to the protosolar value from \cite{2009ARA&A..47..481A}. Rotation is neglected.
% Evolutionary models are built from hydrostatic models computed through a series of 1-bar temperatures $\tone$, calculating the timesteps explicitly from the energy equation
% \begin{equation}\label{eq.energy}
  % \frac{dL}{dm}=-T\frac{ds}{dt}
% \end{equation}
% using forward finite differences. Deep temperatures are obtained by explicitly integrating $\grad$, nominally assumed to be adiabatic but superadiabatic in helium gradient regions if $\rrho>0$, as described above.

\subsection{Rainout and convection}\label{s.rainout_convection}
The internal flux in the models presented here is assumed to be carried purely by convection such that $\nabla=\grada$ to a good approximation, except in cases where emergent helium gradients may partially stabilize the fluid against convection. (Here $\nabla\equiv\frac{{\rm d}\ln T}{{\rm d}\ln P}$ is the temperature gradient in the model and $\grada\equiv\left(\frac{\partial\ln T}{\partial\ln P}\right)_s$ is the adiabatic gradient.) In such cases the double-diffusive instability may operate, and the ensuing nonlinear motions may establish superadiabatic temperature gradients $\grad>\grada$ consistent with a Schwarzschild-unstable, Ledoux-stable configuration.

The overall heat and compositional flux through such a configuration are sensitive to the microscopic diffusivities of heat and solute via the Prandtl number $\pr=\nu/\kt$ and diffusivity ratio $\kmu/\kt$, where $\nu$ is the kinematic viscosity, $\kt$ is the thermal diffusivity set by electron conductions, and $\kmu$ is the diffusivity of solute. However, given the likelihood that excess helium can aggregate by diffusion and rain out of the mixture quickly compared to convection timescales \citep{1973ApJ...181L..83S,ss77b}, non-diffusive processes play an important role and it is not clear whether significant growth rates are achievable by overstable modes. In fact, if rainout of excess helium is fast even compared to the fluid's buoyancy frequency, then an adiabatically perturbed fluid parcel is perennially in equilibrium with its surroundings in terms of solute abundance. This lack of helium contrast between the parcel and its environment means that the buoyancy is no longer affected by helium gradients, and the condition for convective instability reduces back to the Schwarzschild criterion so that $\grad\approx\grada$ should be expected.

In lieu of any detailed understanding of how helium immiscibility affects the double-diffusive instability and associated secondary instabilities like layer formation \citep{2012ApJ...750...61M,2013ApJ...768..157W}, we apply the same simple, generic model as in \cite{2016ApJ...832..113M}. The temperature gradient is allowed to take on superadiabatic values in helium gradient regions, the value of the superadiabaticity assumed to be proportional to the magnitude of the helium gradient there following
\begin{equation}\label{eq.rrho}
  \grad-\grada=\rrho B
\end{equation}
where the ``density ratio'' $\rrho$ (also labeled $R_0$ in the literature) is simply taken as a constant, introducing a free parameter in the model. Here $B$ is the so-called Ledoux term accounting for the effect of composition gradients on the buoyancy frequency \citep[e.g.,][]{1989nos..book.....U}.

Regardless of their helium distributions, stabilizing $Z$ gradients may be a general feature of gas giants as a result of the core accretion process \citep{2017ApJ...840L...4H} or core miscibility \citep{2012PhRvL.108k1101W,2012ApJ...745...54W}, although outcomes in the latter vary widely depending on the stratification of the core boundary \citep{2017ApJ...849...24M}. If these heavy element gradients do exist, they generally dramatically affect the cooling history of the gas giants \citep{2013NatGe...6..347L,2015ApJ...803...32V,2016ApJ...829..118V,2018A&A...610L..14V} and thus deserve close attention. Nonetheless, for conceptual simplicity these models make the strong assumption that the heavy elements are distributed trivially into a distinct $Z=1$ core of chosen mass $\mc$ and an envelope with uniform $Z=Z_1$. In this case the only continuous composition gradients are in the helium mass fraction $Y$, and the term $B$ of Equation~\ref{eq.rrho} reduces to
\begin{equation}\label{eq.b}
  B=\frac{\chi_\rho}{\chi_T}\left(\frac{\partial\ln\rho}{\partial\ln Y}\right)_{P, T}\grad_Y
\end{equation}
where
\begin{equation}\label{eq.chirho_chit}
\chi_\rho=\left(\frac{\partial\ln P}{\partial\ln\rho}\right)_{T,Y}\quad{\rm and}\quad\chi_T=\left(\frac{\partial\ln P}{\partial\ln T}\right)_{P,Y}
\end{equation}
and $\grad_Y=\frac{{\rm d}\ln Y}{{\rm d}\ln P}$ is the true $Y$ gradient in the model.

\subsection{Model atmospheres and Jupiter's Bond albedo}\label{s.atmospheres}
A fundamental input for a planetary evolution model is the surface boundary condition that sets how quickly the planet can cool.
While the total emitted power from the jovian planets is fairly well constrained \citep{2010JGRE..11511002L,2012JGRE..11711002L}, it is more subtle to determine what fraction is emerging from the planet's deep interior (the intrinsic flux derived from thermal energy, contraction and interior processes) as opposed to being re-radiated from absorbed stellar light. Measuring the latter requires broad coverage in phase angle, a requirement that was met in \cassini ISS/VIMS observations during the spacecraft's flyby of Jupiter in 2000-2001. \cite{2018NatCo...9.3709L} analyzed these data to arrive at a new measurement of $A=0.503\pm0.012$ for Jupiter's Bond albedo, a significant departure from the \voyager-era result of $A=0.343\pm0.032$ \citep{1981JGR....86.8705H}. As described by \cite{2018NatCo...9.3709L}, the new larger reflectivity obtained for Jupiter's atmosphere indicates that the planet's intrinsic flux is substantially higher than previously believed.

We apply the model atmospheres of \cite{fortney2011} for Jupiter and Saturn. These models assume no particular Bond albedo, instead solving for a self-consistent radiative-convective equilibrium accounting for the absorption of stellar flux. These models are consistent with \voyager estimates for each planet's Bond albedo, and thus the intrinsic flux they predict for Jupiter at its present-day surface gravity and $\teff$ falls below the recent \cassini measurement. In order to apply a more realistic surface condition for Jupiter, we adjust the $\tint$ column of the \cite{fortney2011} Jupiter tables to instead be consistent with the Bond albedo reported by \cite{2018NatCo...9.3709L}. In particular, we recompute $T_{\rm int}$ from
\begin{equation}
  \teff^4=\tint^4+\teq^4
\end{equation}
where $T_{\rm eff}$ is as given in the tables and $T_{\rm eq}=102.5\ {\rm K}$ is the value implied by $A=0.503$ for the solar flux received at Jupiter's semimajor axis. The predictions of the model atmospheres modified in this way deviate from their tabulated values only at relatively late times when the absorbed solar flux (proportional to $\teq^4$) becomes significant compared to the intrinsic flux ($\propto\tint^4$).

The models in this work assume that the solar luminosity has increased linearly as a function of time from $0.7\,L_\odot$ at $0\ {\rm Gyr}$ to $1.0\,L_\odot$ at $4.56\ {\rm Gyr}$. To this end $\tint$ and $\teff$ are evaluated for both of these bracketing values for the solar flux as tabulated by \cite{fortney2011}, and interpolation in age between the two provides the $\tint$ and $\teff$ adopted in a given timestep. The sole exception is in Figure~\ref{fig.homog} described below, where evolutionary curves assuming a static solar luminosity at $1.0\,L_\odot$ are shown for comparison. This simplification tends to overestimate the cooling time for Jupiter or Saturn by at least $10^8$ yr compared to the more realistic case accounting for the sun's evolving luminosity.

\subsection{Expectations from simpler evolution models}
\begin{figure*}[t]
    \begin{center}
        \includegraphics[width=\textwidth]{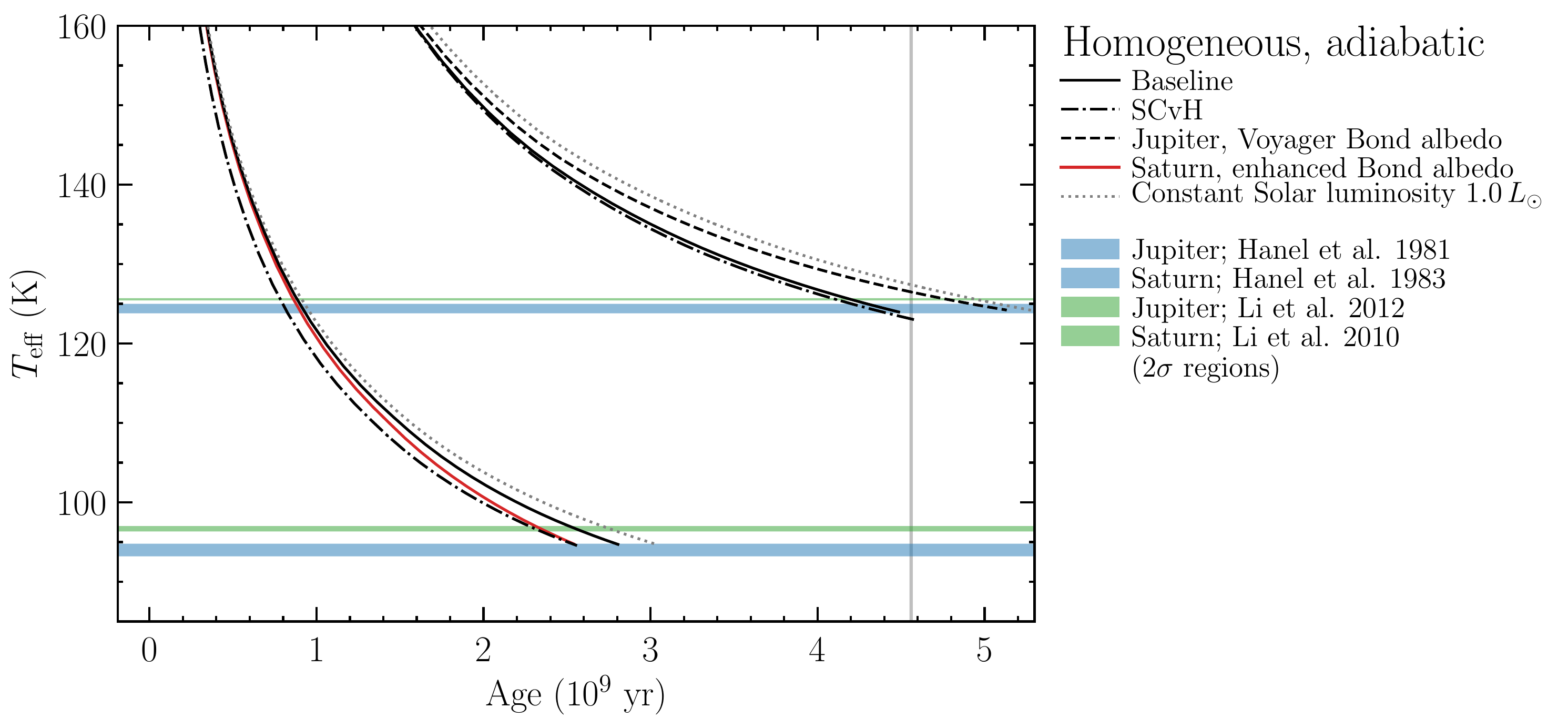} % {../figures/homog_cooling_curves_teff}
        \caption{
        The importance of the assumed hydrogen EOS and surface boundary condition for Jupiter and Saturn cooling times. ``Baseline'' uses the up-to-date Bond albedos and the \cite{2013ApJ...774..148M} hydrogen-helium EOS combined with \cite{1995ApJS...99..713S} helium as described in the text. Shaded regions mark the observed effective temperatures and the solar age $\tsun=4.56\ {\rm Gyr}$.
        }
        \label{fig.homog}
    \end{center}
\end{figure*}

To illustrate the overall influence of the atmospheric surface condition on cooling times, we show in Figure~\ref{fig.homog} baseline cooling curves for Jupiter and Saturn assuming homogeneous, adiabatic interiors. The \voyager \citep{1981JGR....86.8705H} and \cassini \citep{2018NatCo...9.3709L} determinations of Jupiter's Bond albedo are compared. Analysis of existing \cassini data may also reveal a higher albedo for Saturn, and thus for that case we compare the \voyager \citep{1983Icar...53..262H} Bond albedo is to a hypothetical higher value of $A=0.5$; this scenario will be revisited in detail in Section~\ref{s.saturn_results} below. Two EOSs for hydrogen \citep{2013ApJ...774..148M,1995ApJS...99..713S} are also compared. The assumed EOS has only a modest effect on Jupiter's cooling time, whereas the updated surface condition accelerates the time for Jupiter to cool to its observed $\teff$ by a significant few times $10^8$ yr. For Saturn, both the EOS and the surface condition significantly modify the cooling time, but in any case homogeneous models fail to explain Saturn's heat flow at the solar age $t=\tsun=4.56\ {\rm Gyr}$, recovering a well-known result \citep{1977Icar...30..111P,1980Icar...42..358G,fortney2011}.

It is significant that Jupiter's revised albedo brings cooling times for homogeneous models short of 4.56 Gyr, because it means that some amount of additional luminosity from differentiation of the planet's chemical components can be straightforwardly accommodated. Jupiter and Saturn now both require an extra luminosity source, and given our expectations for the relative amounts of helium lost from each planet's molecular envelope (Section~\ref{s.phase}), it appears that helium immiscibility may provide one natural explanation. The success of this scenario in explaining the observed heat flow in both planets is assessed in detail in Section~\ref{s.results} below.

\subsection{Parameter estimation}\label{s.mcmc}
The evolution models here contain four free parameters. The core mass $\mc$ and envelope heavy element abundance $Z_1$ control the distribution of heavy elements. As described in Section~\ref{s.dtphase}, the phase diagram temperature offset $\dt$ controls the temperatures of the hydrogen-helium phase curves and thus dictates the overall amount of helium differentiation. Finally the density ratio $\rrho$ sets the superadiabaticity of the temperature profile in regions with continuous helium gradients, providing an additional degree of freedom in setting the rate of cooling from the planet's surface by limiting the flux emerging from the metallic interior \citep[e.g.,][]{2015MNRAS.447.3422N,2016ApJ...832..113M}.

We estimate these parameters independently for each planet by starting with the planets' observed mean radii and effective temperatures at the solar age 4.56 Gyr, applying Bayes' theorem assuming that the likelihood of the data given a model follows a three-dimensional normal distribution with trivial covariance, and sampling from the posterior probability distribution using the ensemble Markov chain Monte Carlo sampler \texttt{emcee} \citep{2013PSP..125..306F}.  Uniform priors are assigned to each of the four parameters within the ranges $0<\mc<30$, $0<Z_1<0.5$, $|\dt|<2,000\ {\rm K}$, and $0\leq\rrho<1$
except in cases where otherwise noted. Samples are judged to be converged based on inspection of the posterior distributions and the individual traces of each walker in the sampler. The samples described below consist of between 30,000 and 60,000 evolutionary models each.

\begin{deluxetable*}{rccl}
\tabletypesize{\footnotesize}
\tablecolumns{4}
\tablewidth{1.0\columnwidth}
\tablecaption{\label{t.data}Jupiter and Saturn evolutionary constraints}
\tablehead{
\colhead{Quantity} & \colhead{Jupiter} & \colhead{Saturn} & \colhead{Reference}
}
\startdata
  $\teff\ ({\rm K})$\tablenotemark{a} & $125.57\pm0.07$ & $96.67\pm0.17$ & \cite{2012JGRE..11711002L,2010JGRE..11511002L} \\
  $R\ ({\rm km})$\tablenotemark{b} & $69,911\pm70$ & $58,232\pm58$ & \cite{2007CeMDA..98..155S} \\
  ${\rm Age}\ (10^9\ {\rm yr})$ & \multicolumn{2}{c}{4.56$\pm$0.10} & \cite{2012Sci...338..651C} \\
  $Y_1/(X_1+Y_1)$ & $0.238\pm0.005$ & -- & \cite{1998JGR...10322815V}
\enddata
\tablenotetext{a}{Mean value used as the condition for model-data comparison rather than fit; see Section~\ref{s.mcmc}.}
\tablenotetext{b}{Errors are inflated to $10^{-3}$ times the mean value, about ten times the true volumetric radius uncertainty.}
% \tablecomments{Comments}
\end{deluxetable*}

For Jupiter models, the likelihood includes an additional term comparing the \galileo probe interferometric measurement of Jupiter's atmospheric helium abundance \cite{1998JGR...10322815V} to the abundance in the well-mixed molecular envelope of the models. Because this interferometric measurement (along with the variety of estimates obtained for Saturn from thermal emission, occultation, and limb scan data) is ultimately sensitive to the He/H$_2$ mixing ratio, comparisons are made in terms of the helium mass fraction relative to the hydrogen-helium mixture:
\begin{equation}
  \frac{Y_1}{X_1+Y_1}=\frac{Y_1}{1-Z_1}.\label{eq.y_xy}
\end{equation}
Here and in what follows $X_1$, $Y_1$ and $Z_1$ denote the true mass fractions of hydrogen, helium, and water respectively in the well-mixed molecular envelope of our model Jupiters and Saturns.

In reality, a slightly less intuitive approach than this is used because the 1-bar temperature $\tone$ is the fundamental independent variable rather than time $t$. In particular, our method can guarantee that a given model eventually cools to the correct $\teff$, but there is no guarantee that the same model will reach the solar age $t=\tsun$ before it cools out of the regime covered by the model atmospheres. Therefore, although $\teff$ is in fact the relatively uncertain quantity and the solar age the relatively certain one, we instead cool models to their observed $\teff$ and then treat their age as the uncertain data point, distributed normally about 4.56 Gyr with a somewhat arbitrary standard deviation equal to 0.10 Gyr. This approach has the advantage of bestowing even poorly fitting models with meaningful likelihoods, whereas if model-data comparisons were always made at 4.56 Gyr it would not be clear what to do with a Saturn model that cooled in, e.g., 4.4 Gyr. In practice all models are cooled through the planet's $\teff$, and quantities compared to data ($R$, $Y_1$, age) are linearly interpolated within the timestep spanning that $\teff$.

The data used as constraints for Jupiter and Saturn's thermal evolution are summarized in Table~\ref{t.data}.

\section{Results}\label{s.results}
As described in Section~\ref{s.mcmc} above, the evolutionary models here have four tunable parameters that are sampled using Bayesian parameter estimation from the observed effective temperatures and radii at the solar age. In the case of Jupiter the \galileo helium constraint is used as an additional constraint.
In what follows we devote our attention to the parameters $\dt$ and $\rrho$ pertaining directly to the helium distributions and thermal histories of Jupiter and Saturn, addressing each planet in turn. Because our evolutionary models forgo any detailed calculations of rotation, oblateness, and the associated zonal gravity harmonics, we find as expected that $\mc$ and $Z_1$ are extremely degenerate, essentially unconstrained so long as the total heavy element content is sufficient to fit each planet's mean radius at the solar age. As a result the two parameters are strongly anticorrelated.

\subsection{Jupiter}\label{s.jupiter_results}
\begin{figure}[t]
    \begin{center}
        \includegraphics[width=1.0\columnwidth]{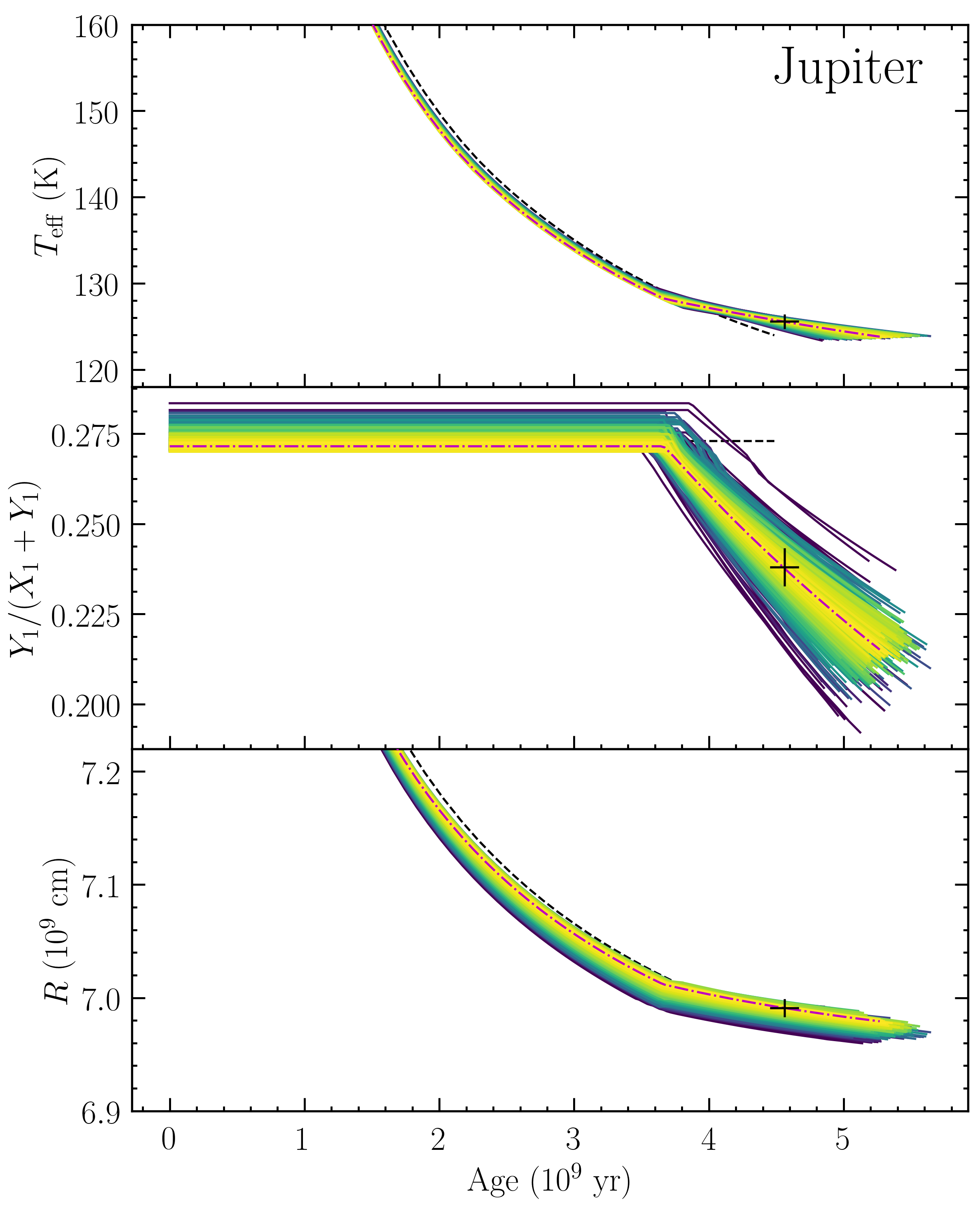} % {../figures/jup_tracks.png}
        \caption{
        Jupiter's evolution with instantaneous helium rainout following the \sch phase diagram. Tracks are colored by their log posterior probability with more likely models appearing yellow and progressively less likely samples appearing green to blue to purple. Black crosses signify the data summarized in Table~\ref{t.data}; the observed $\teff$ is displayed with $10\sigma$ errors for clarity. The most likely individual model is shown as the dot-dashed magenta curve.
        }
        \label{fig.jup_tracks}
    \end{center}
\end{figure}

The evolutionary paths we obtain for our baseline Jupiter model are shown in Figure~\ref{fig.jup_tracks}, which illustrates the basic success of this model in matching Jupiter's effective temperature, mean radius, and atmospheric helium content at the solar age. (Here ``baseline" is used to distinguish from the alternative case where the \voyager value of Jupiter's Bond albedo \citep{1981JGR....86.8705H} is used; this alternative case is compared below.) The intrinsic spread in these cooling curves and in all those that follow stem from the uncertainties in the observational constraints summarized in Table~\ref{t.data}; these uncertainties translate directly to variance in the posterior distributions of the model parameters $\mc$, $Z_1$, $\rrho$, and $\dt$.

\begin{figure}[t]
    \begin{center}
        \includegraphics[width=1.0\columnwidth]{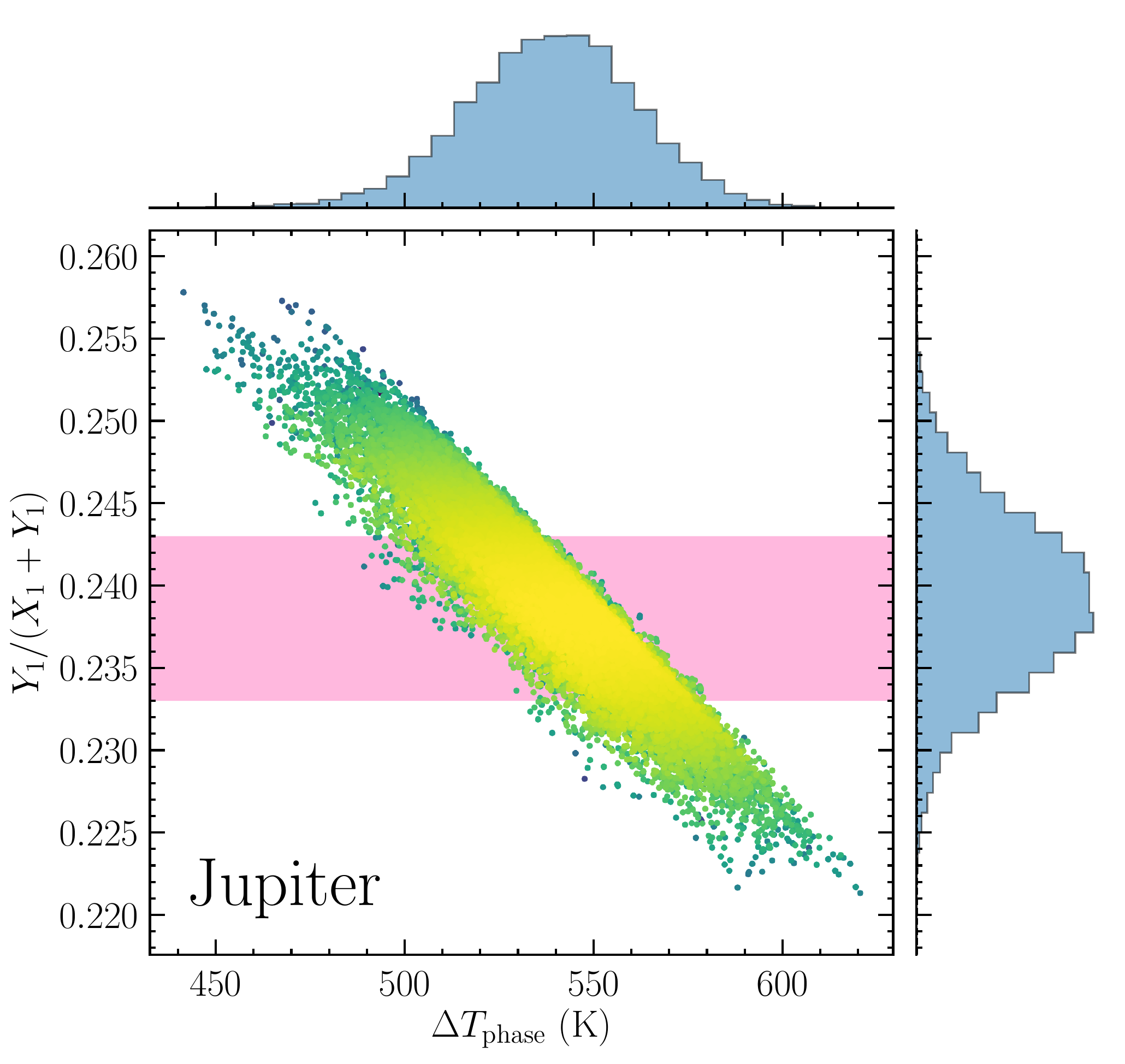} % {../figures/jup_dt_y1}
        \caption{
        Prediction for the temperature of the true phase diagram (relative to the unperturbed \sch diagram) based on the atmospheric helium content of Jupiter. The shaded band represents the \galileo probe interferometer measurement \citep{1998JGR...10322815V}, and the color of model points maps to log posterior probability as in Figure~\ref{fig.jup_tracks}. Shown is the baseline (superadiabatic) Jupiter sample, but other Jupiter cases (adiabatic; low albedo) yield virtually identical distributions on this diagram.
        }
        \label{fig.jup_dt_y1}
    \end{center}
\end{figure}

The first conclusion we can reach based on the solutions obtained for Jupiter is that our model would rule out a phase diagram as cold as \sch, which unperturbed ($\dt=0$) leads to no differentiation of helium such that the model overestimates Jupiter's atmospheric helium abundance relative to the \galileo measurement (Figure~\ref{fig.jup_dt_y1}). This confirms expectations from of Section~\ref{s.dtphase} and the findings of \cite{2018PhRvL.120k5703S}, who predicted based on a comparison to reference adiabats that helium immiscibility is marginal or absent in Jupiter. We find that translating the \sch phase curves to higher temperatures at $\dt=(539\pm23)\ {\rm K}$ instead gives excellent agreement. Furthermore $\dt$ shows very weak covariance with other parameters, being constrained almost entirely by the \galileo measurement.

\begin{figure}[t]
    \begin{center}
        \includegraphics[width=1.0\columnwidth]{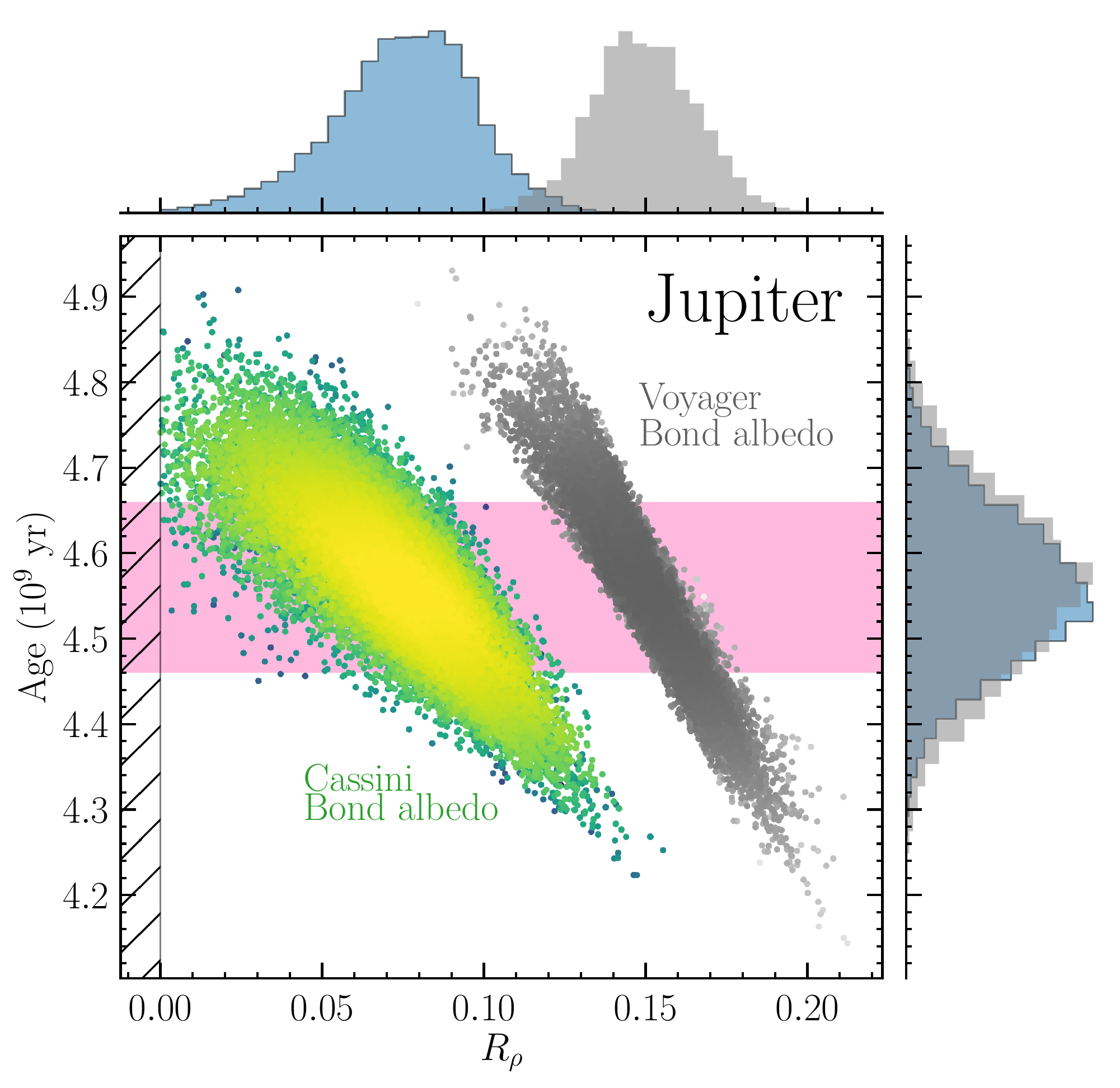} % {../figures/jup_rrho_age}
        \caption{
        Prediction for the fractional superadiabaticity (Equation~\ref{eq.rrho}) in Jupiter's helium gradient region. Two different assumptions regarding the atmospheric boundary condition are compared; one is the recent \cassini Bond albedo measurement from \citet[][colorful points and blue histograms]{2018NatCo...9.3709L} and the other is the \voyager measurement from \citet[][grey points and histograms]{1981JGR....86.8705H}. The shaded band represents the imposed age constraint.
        }
        \label{fig.jup_rrho_age}
    \end{center}
\end{figure}

As discussed above, any superadiabatic regions associated with helium gradients bear on cooling times directly, because they can generally trap heat in the deep interior and cause the molecular envelope to cool relatively quickly. Larger $\rrho$ can thus generally mitigate the cooling time extension offered by helium differentiation, although speaking generally, the process can become complicated by the feedback between the temperature profile and the equilibrium helium distribution providing the stratification. Nonetheless, as in \cite{2016ApJ...832..113M}, the simple picture just described is the general behavior observed in our Jupiter models after helium begins differentiating. The relationship between superadiabaticity $\rrho$ and cooling time is illustrated in Figure~\ref{fig.jup_rrho_age}, where results for the baseline model are compared with those assuming the older \voyager Bond albedo from \cite{1981JGR....86.8705H}. The comparison reveals that Jupiter's updated albedo allows excellent fits at substantially lower superadiabaticity, although perfect adiabaticity appears to be ruled out. This is qualitatively consistent with the findings in \cite{2016ApJ...832..113M}, where once the Bond albedo was treated as a free parameter, solutions with larger albedos (lower $\teq$) and interiors closer to adiabatic were recovered. The quantitative results obtained there were different from the present results because of the more strongly restrictive prior assumed for $\rrho$ there (see Section 2.4 of that work), and to a lesser degree the older, likely less realistic EOS and phase diagram used in those calculations.

\subsection{Saturn}\label{s.saturn_results}
\begin{figure}[t]
    \begin{center}
        \includegraphics[width=1.0\columnwidth]{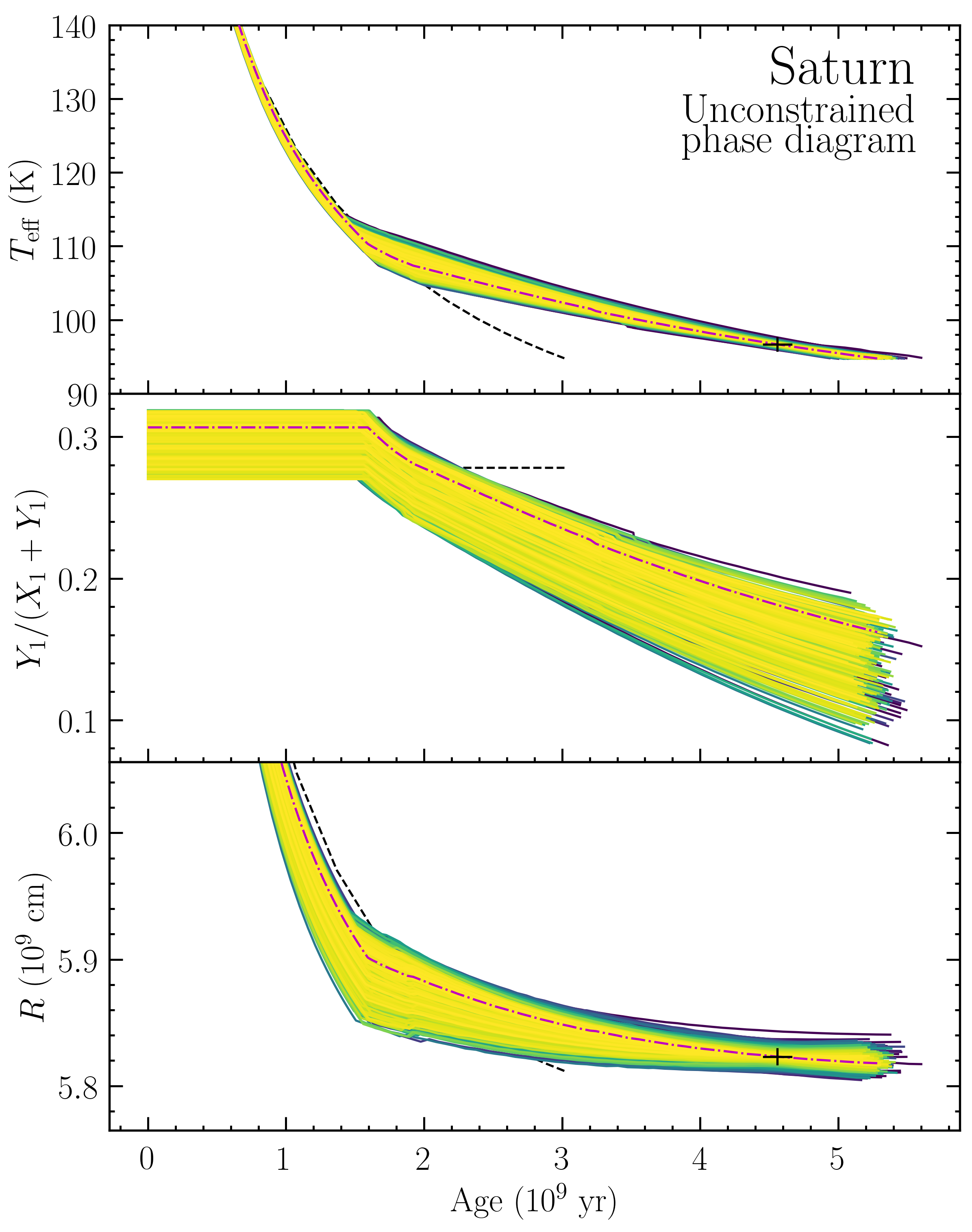} % {../../sat_lsun/tracks.png}
        \caption{
        As in Figure~\ref{fig.jup_tracks}, but for Saturn. This is our baseline (most general) Saturn sample with a uniform prior on $\dt$. An ostensibly good individual model is shown in dot-dashed magenta. None of these evolutionary tracks are favored because they require cold phase diagrams that would overpredict Jupiter's observed helium depletion; see Figure~\ref{fig.sat_dt_age} and discussion in the text.
        % Individual histories are plotted for two ostensibly good solutions: the most likely model with $\dt<550\ {\rm K}$ is shown in the magenta dot-dashed curve, and the most likely model with $\dt>550\ {\rm K}$ is shown in the grey dashed curve. The latter case exhausts the helium in its molecular envelope entirely at $t\approx4.1\ {\rm Gyr}$; this type of evolution is not favored because it requires a phase diagram incompatible with Jupiter's observed helium depletion.
        }
        \label{fig.sat_tracks}
    \end{center}
\end{figure}

\begin{figure}[t]
    \begin{center}
        \includegraphics[width=1.0\columnwidth]{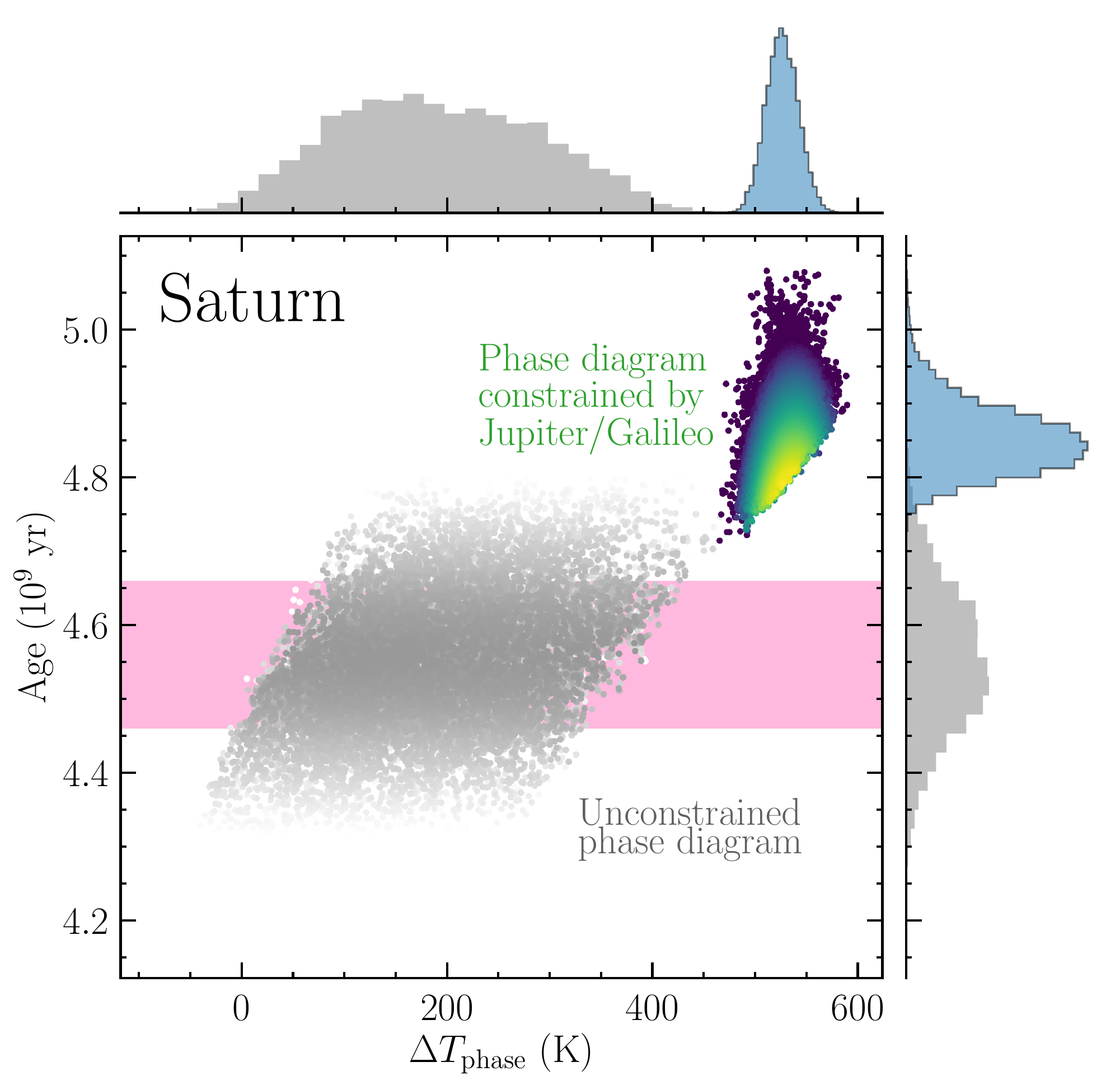} % {../../figures/sat_dt_age}
        \caption{
        The influence of $\dt$ on Saturn's cooling time. The grey distribution represents our initial, most general Saturn sample, tracks from which are shown in Figure~\ref{fig.sat_tracks}. This distribution is disfavored because the resulting phase diagrams ($\dt\lesssim400\ {\rm K}$) are inconsistent with our Jupiter models which require $\dt=(539\pm23)\ {\rm K}$. The colorful distribution (with blue marginalized posteriors at right and top) is our sample obtained by imposing $\dt=(539\pm23)\ {\rm K}$ as a Gaussian prior probability representing the family of allowable phase diagrams based the Jupiter models that reproduce the \galileo helium measurement. This second, more tightly constrained distribution of Saturn models is also disfavored because it produces Saturn cooling times significantly longer than the solar age.
        % As in Figure~\ref{fig.sat_dt_y1}, but this time showing the influence of $\dt$ on Saturn's cooling time. The non-monotone relationship for the sample with no constraint on the phase diagram is caused by the models with $\dt\gtrsim550\ {\rm K}$ losing the helium from their molecular envelopes entirely before the solar age, evolving on a rapid cooling track from that time.
        }
        \label{fig.sat_dt_age}
    \end{center}
\end{figure}

We find an abundance of Saturn models that successfully explain Saturn's heat flow at the solar age.
A random subset of these cooling tracks are presented in Figure~\ref{fig.sat_tracks}. Because no definitive constraint is available for Saturn's atmospheric helium content at the solar age, and a uniform prior probability is assigned to $\dt$, this sample explores a wide variety of phase diagrams via $\dt$. This manifests in the large spread of evolutionary paths for $Y_1/(X_1+Y_1)$ in the middle panel of Figure~\ref{fig.sat_tracks} and the quite broad posterior $\dt$ distribution shown in Figure~\ref{fig.sat_dt_age} with the label ``unconstrained phase diagram.''
As is the case for Jupiter, {most} good solutions for Saturn require that \sch be translated to warmer temperatures, this time driven by Saturn's luminosity constraint: taken at face value, the \sch phase curves {($\dt=0\ {\rm K}$)} predict insufficient differentiation in Saturn to {robustly} provide the planet's observed luminosity.
In the other direction, {phase curves with $\dt\gtrsim400\ {\rm K}$} lead to pronounced differentiation such that Saturn is generally overluminous at the solar age{, producing a dearth of solutions there}.

\subsubsection{A better-informed phase diagram}\label{s.sat_dt}
Although successful in terms of Saturn's luminosity and radius constraints, the Saturn models described so far present a major problem in that they require phase diagrams that at $\dt=0-400\ {\rm K}$ are not consistent with the $\dt=(539\pm23)\ {\rm K}$ necessary to explain Jupiter's observed atmospheric helium abundance. The same fundamental physics is at work within both planets; in the interest of applying a consistent physical model to both, we carry out a new calculation for Saturn that incorporates our belief about the true phase diagram as informed by the Jupiter models that satisfy the \galileo measurement (see Section~\ref{s.jupiter_results} and Figure~\ref{fig.jup_dt_y1}).
This updated Saturn sample imposes a prior likelihood for $\dt$ proportional to the marginalized posterior distribution obtained for Jupiter and driven by the \galileo probe measurement of Jupiter's atmospheric helium abundance. This distribution is fit well by a normal distribution with mean $539\ {\rm K}$ and standard deviation $23\ {\rm K}$ (see Section~\ref{s.jupiter_results}).

This better informed phase diagram leads to a more restrictive sample of Saturns. Figure~\ref{fig.sat_dt_age} reveals that the strong constraint imposed on $\dt$ leads to Saturn models that differentiate substantially enough that their cooling time is unrealistically long in all cases. The fact that this sample fails to meet at least one of the observational constraints is not unexpected, since the original Saturn sample (grey distributions in Figure~\ref{fig.sat_dt_age}) yielded no probability density in the neighborhood of $\dt=(539\pm23)\ {\rm K}$. Because this basic observational constraint is violated, these Saturn models also appear to be inadequate.

\subsubsection{Saturn's Bond albedo}
Faced with the failure of the Saturn models obtained so far to simultaneously provide (1) physical consistency with the Jupiter evolution models of Section~\ref{s.jupiter_results} vis-à-vis the phase diagram and (2) a satisfactory fit to the basic observables for Saturn, we finally explore the possibility that Saturn's Bond albedo deviates from the \cite{1983Icar...53..262H} value $A=0.342\pm0.030$ assumed in our models so far.

A revised, and larger, Bond albedo for Saturn is actually quite likely given already published data.  It is well known that the reflection spectra of both Jupiter and Saturn are dominated by scattering from ammonia clouds, and absorption due to gaseous methane and a methane-derived photochemical haze.  Of particular interest, \cite{1994Icar..111..174K} have previously shown that Saturn's optical geometric albedo (as measured from the Earth, seen at full phase) is slightly larger that Jupiter's in a wide optical bandpass (300 to 1000 nm).  Since both Saturn and Jupiter had \voyager era Bond albedo derivations of $\sim$~0.34, this suggests that Saturn could see a similar revision to $\sim$~0.5 once \cassini data are analyzed.  An additional supporting point of view comes from the modeling work from \cite{2010ApJ...724..189C}.  They found that Jupiter- and Saturn-like giant planet atmosphere models that yielded wavelength-dependent geometric albedos similar to that of the \cite{1994Icar..111..174K} data yielded Bond albedos of $\sim$~0.5, not $\sim$~0.34.

% \cite{2018NatCo...9.3709L} demonstrated that relative to the previous best estimate by \cite{1981JGR....86.8705H}, the more complete wavelength and phase angle coverage afforded by a \cassini flyby led to a Bond albedo for Jupiter larger than Hanel et al.'s measurement by more than 13 times the latter's estimated systematic uncertainty. While no analogous estimate has yet been made for Saturn from \cassini data, it seems plausible that the same unquantified systematic biases may have led the similar measurement by \cite{1983Icar...53..262H} to understimate Saturn's Bond albedo.
It is with this in mind that we consider a final ``enhanced Bond albedo" case for Saturn, adopting $A=0.5$ as a representative value close to the estimate for Jupiter. Figure~\ref{fig.homog} shows that this scenario leads to a cooling time for a homogeneous, adiabatic Saturn more than $0.2$ Gyr shorter than that for the baseline homogeneous model. This more reflective atmospheric boundary condition thus tends to mitigate the long cooling times found so far for Saturn when imposing the Jupiter/\galileo phase diagram constraint.

\begin{figure}[t]
    \begin{center}
        \includegraphics[width=1.0\columnwidth]{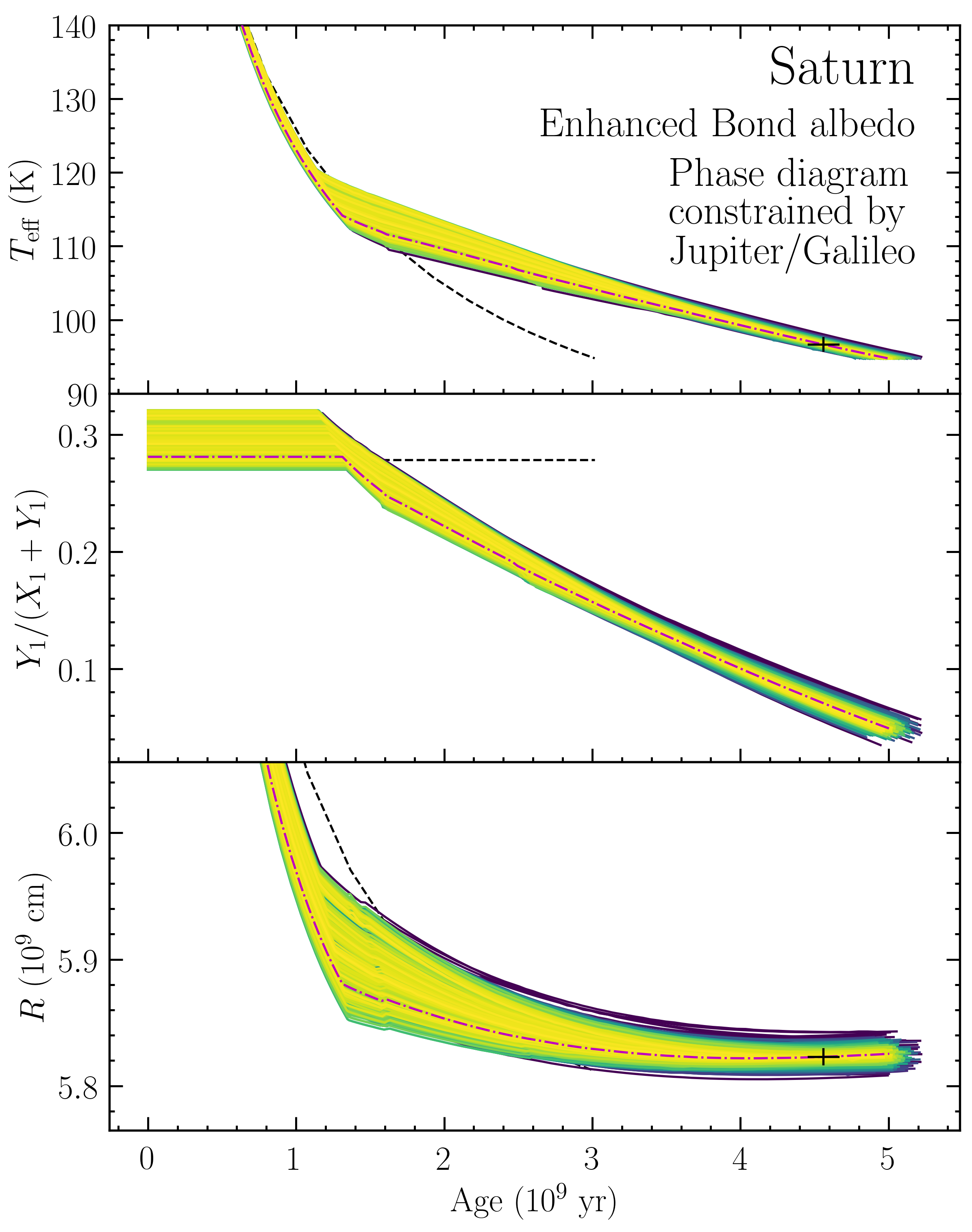} % {../../sat_dt_lsun_high_albedo/tracks.png}
        \caption{
        As in Figure~\ref{fig.sat_tracks}, but applying phase diagrams that satisfy the \galileo measurement of Jupiter's atmospheric helium abundance, and assuming an enhanced Bond albedo $A=0.5$ to allow for the possibility that Saturn's true Bond albedo is significantly larger than the current best estimate by \cite{1983Icar...53..262H}; see discussion in text. This is our favored class of model for Saturn based on goodness of fit and physical consistency with the Jupiter models.
        }
        \label{fig.sat_dt_high_albedo_tracks}
    \end{center}
\end{figure}

\begin{figure}[t]
    \begin{center}
        \includegraphics[width=1.0\columnwidth]{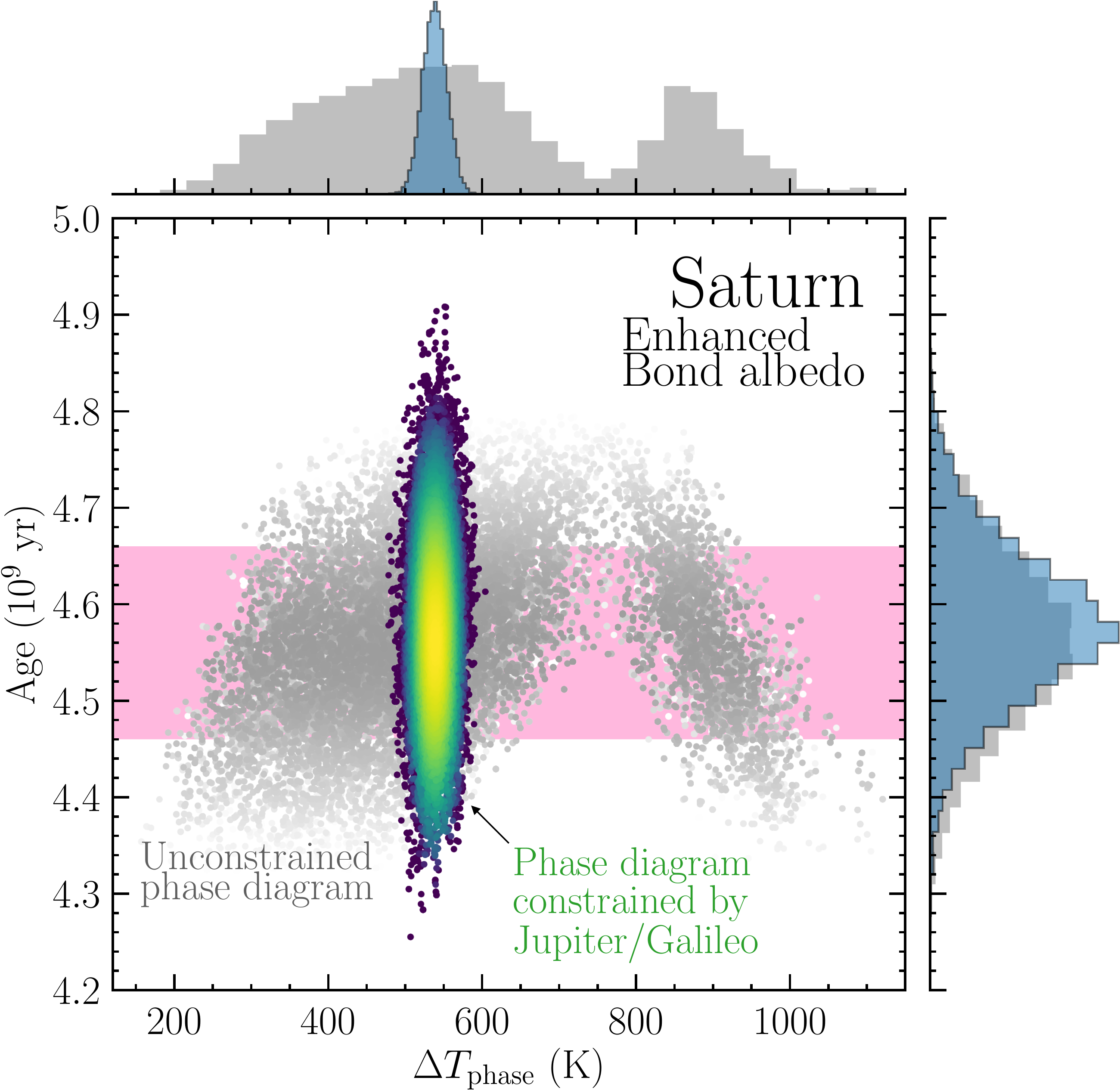} % {../../figures/sat_dt_age_high_albedo_annotate-crop}
        \caption{
        As in Figure~\ref{fig.sat_dt_age}, but for models assuming an enhanced Bond albedo $A=0.5$ for Saturn. The colorful distribution incorporates prior belief for allowable phase diagrams based on our Jupiter results and represents our favored class of Saturn model; the grey distribution neglects this constraint.
        }
        \label{fig.sat_dt_age_high_albedo}
    \end{center}
\end{figure}

\begin{figure}[t]
    \begin{center}
        \includegraphics[width=1.0\columnwidth]{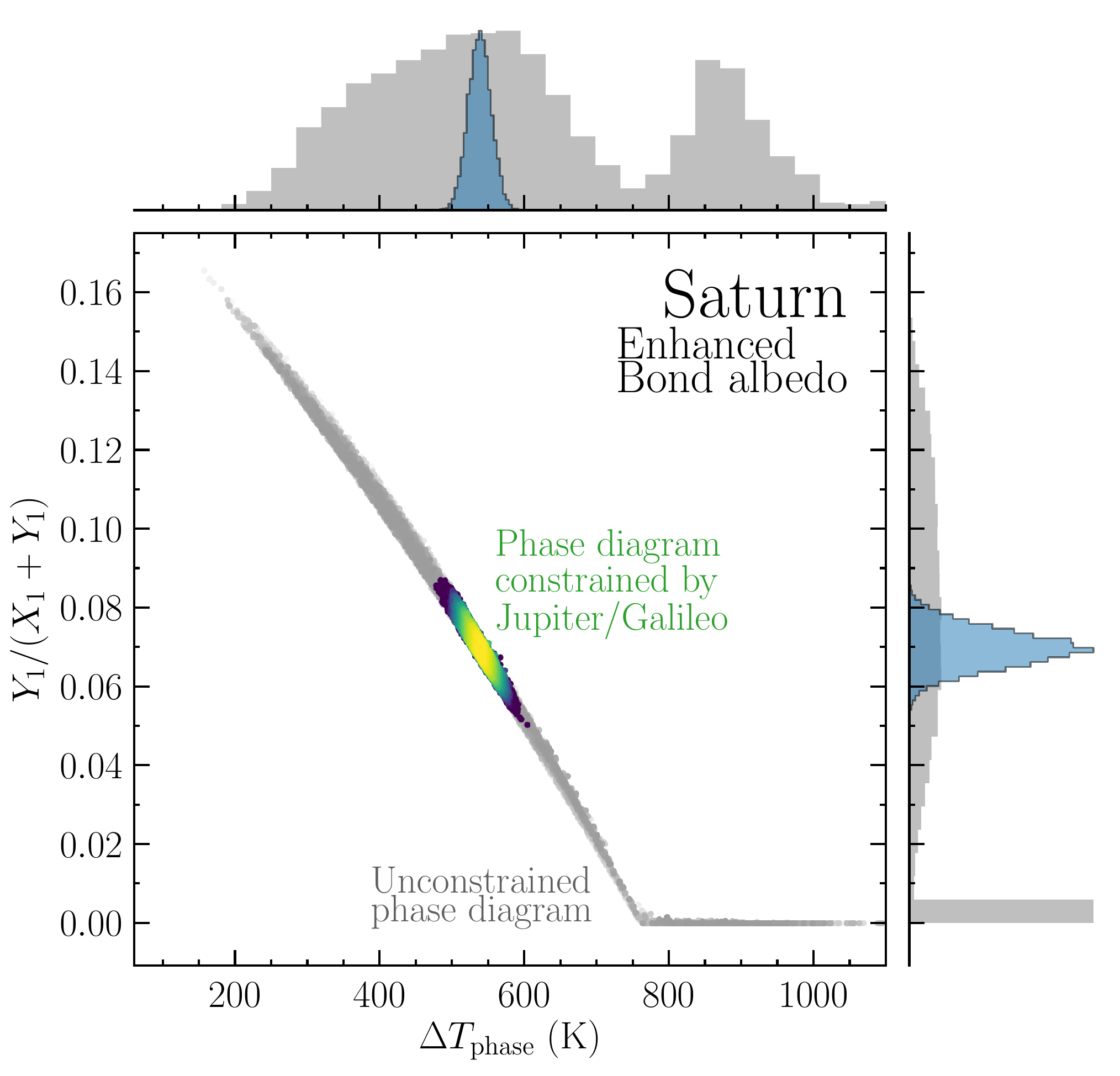} % {../../figures/sat_dt_y1_high_albedo}
        \caption{
        As in Figure~\ref{fig.sat_dt_age_high_albedo}, but showing the atmospheric helium mass fraction $Y_1/(X_1+Y_1)$ predicted for Saturn at the solar age.
        }
        \label{fig.sat_dt_y1_high_albedo}
    \end{center}
\end{figure}

Figure~\ref{fig.sat_dt_high_albedo_tracks} presents the resulting evolutionary tracks, where it is seen that the enhanced albedo leads to models that satisfy all basic constraints for phase diagrams allowed by Jupiter/\galileo. Here the models realize a narrow distribution of values for the atmospheric helium mass fraction compared to the tracks in Figure~\ref{fig.sat_tracks}, a direct consequence of incorporating the Jupiter/\galileo phase diagram prior. This distribution will be presented in Section~\ref{s.sat_y1} below. The relationship between $\dt$ and cooling time for these models is shown in Figure~\ref{fig.sat_dt_age_high_albedo}; the effect of $\dt$ on predicted $Y_1$ at the solar age is shown in Figure~\ref{fig.sat_dt_y1_high_albedo}. As in Figure~\ref{fig.sat_dt_age}, these figures both show models obtained with an unconstrained phase diagram (grey distributions) as a comparison case.

In addition to the posterior distribution of $\dt$ being considerably broader for this unconstrained sample, it also exhibits a distinct bimodality that has been absent in all other classes of model considered here. This bimodality arises because the model cooling times exhibit a strongly non-monotone dependence on $\dt$, as is evident in Figure~\ref{fig.sat_dt_age_high_albedo} and has already been characterized in this context by \citet[][their Figure 10]{2016Icar..267..323P}. The first mode corresponds to phase diagrams hotter than \sch by $200-750\ {\rm K}$, where warmer phase curves \emph{extend} Saturn's cooling time because they lead to more pronounced differentiation in the planet. This behavior has a limit, though, corresponding to phase curves that are so warm that they lead to the \emph{complete} exhaustion of the helium that initially resided in the molecular envelope. This is the case for models beyond $\dt\approx750\ {\rm K}$, where hotter phase diagrams push this exhaustion time for Saturn farther into the past. In this limit the models begin to again undershoot the solar age because an increasing fraction of their time is spent in a final episode of rapid cooling after their differentiation luminosity vanishes. This behavior is exhibited by the second mode visible in Figure~\ref{fig.sat_dt_age}, and also manifests in the abundance of models with final $Y_1=0$ in Figure~\ref{fig.sat_dt_y1_high_albedo}.

By construction the Jupiter/\galileo constraint leads to a narrow distribution in Figures~\ref{fig.sat_dt_age_high_albedo} and \ref{fig.sat_dt_y1_high_albedo}, ruling out the hot phase diagrams that lead to complete exhaustion of helium from Saturn's envelope. As seen in Figure~\ref{fig.sat_dt_y1_high_albedo}, this tight constraint on $\dt$ translates directly to a narrow distribution of predicted atmospheric helium abundance for Saturn at the solar age; this distribution is compared with estimates from observations in Section~\ref{s.sat_y1} below.

We note that among our preferred Saturn models, i.e., those applying phase diagrams constrained by Jupiter/\galileo, it is typical for Saturn's total radius $R$ to be stalled or actually increasing as a function of time at the solar age. This behavior runs counter to the normal expectation that a cool gas giant contracts as it cools, and is a consequence of interior thermal energy being made available by differentiation of helium from hydrogen. This type of radius evolution takes place whether the Saturn interiors are superadiabatic ($\rrho>0$) or adiabatic ($\rrho=0$).

\subsubsection{Need Saturn be superadiabatic?}
\begin{figure}[t]
    \begin{center}
        \includegraphics[width=1.0\columnwidth]{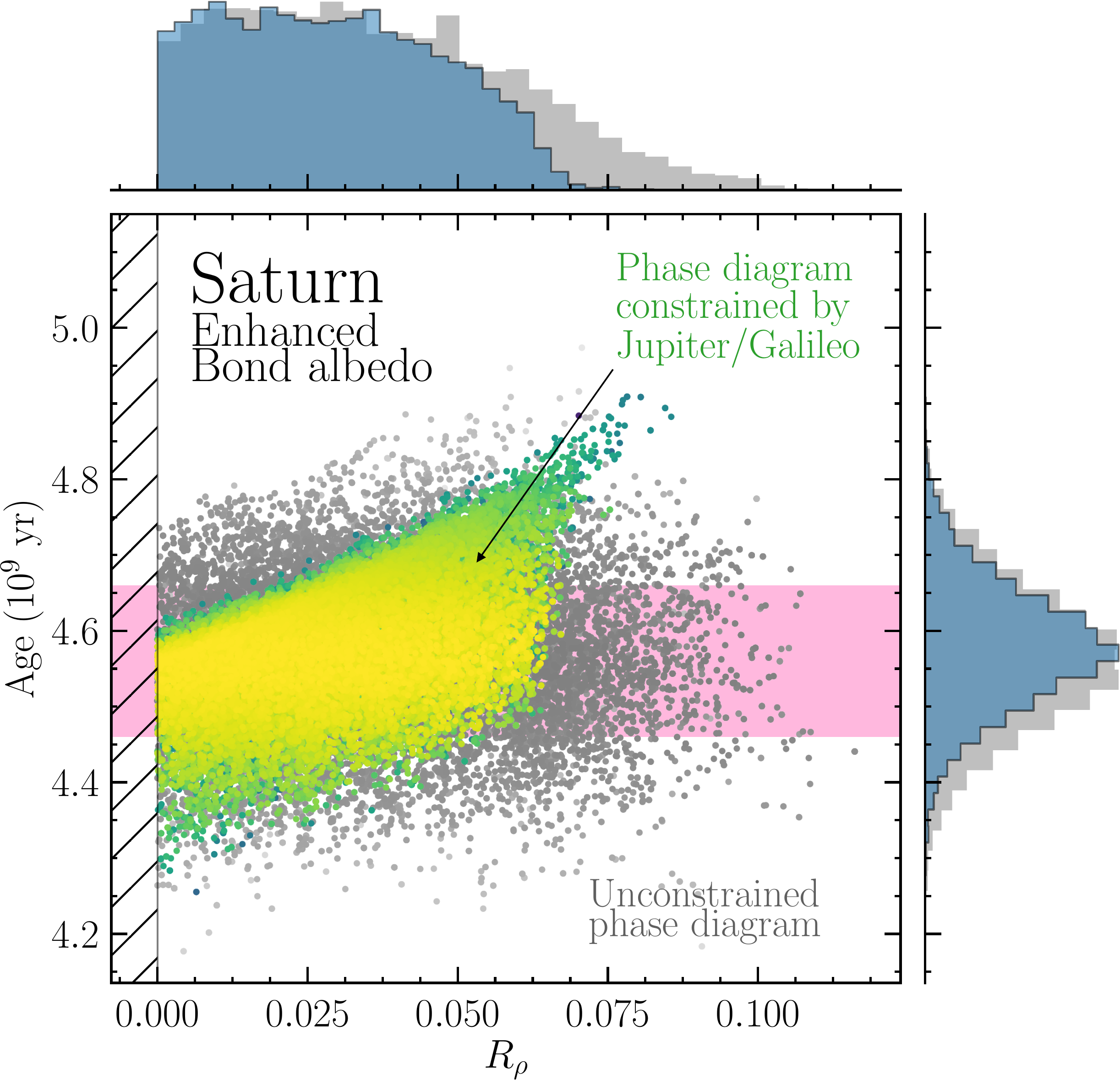} % {../../figures/sat_rrho_age_high_albedo_annotate-crop}
        \caption{
        As in Figure~\ref{fig.sat_dt_age_high_albedo}, but showing the influence of $\rrho$ on Saturn's cooling time. In both cases adiabatic models ($\rrho=0$) achieve good fits.
        }
        \label{fig.sat_rrho_age}
    \end{center}
\end{figure}

Section~\ref{s.jupiter_results} demonstrated that some amount of superadiabatic temperature stratification $\rrho\gtrsim0.05$ is required in Jupiter to recover satisfactory solutions. The distributions in Figure~\ref{fig.sat_rrho_age}, on the other hand, are consistent with $\rrho=0$, leading one to expect that adiabatic models perform roughly as well as the rest in the case of Saturn. To quantify this comparison, we calculate a new sample identical to our favored Saturn model (enhanced Bond albedo; Jupiter/\galileo phase diagram prior) but eliminating the fourth parameter by setting $\rrho=0$ to allow only adiabatically stratified interiors. We then compare the two models in terms of an \cite{1100705} information criterion
\begin{equation}
  {\rm AIC}=2n-2\ln{\mathcal L}_{\rm max},
\end{equation}
where $n$ is the number of free parameters in each sample and $\mathcal L_{\rm max}$ is the maximum likelihood obtained therein.

We find that the adiabatic and superadiabatic samples achieve virtually the same maximum likelihoods ($\ln\mathcal L_{\rm max}=-15.1150$ versus $-15.1141$, respectively) such that the $n=3$ sample yields ${\rm AIC}=36.2282$ while the $n=4$ sample yields ${\rm AIC}=36.2300$. This model comparison suggests that the superadiabatic Saturn models introduce additional model complexity with no return in terms of quality of fit, and thus there is no basis on which to prefer superadiabatic versus adiabatic Saturn thermal evolution models. Possible implications of the success of adiabatic models will be discussed in Section~\ref{s.discussion}.

\subsubsection{Saturn's atmospheric helium content}\label{s.sat_y1}
\begin{figure}[t]
    \begin{center}
        \includegraphics[width=1.0\columnwidth]{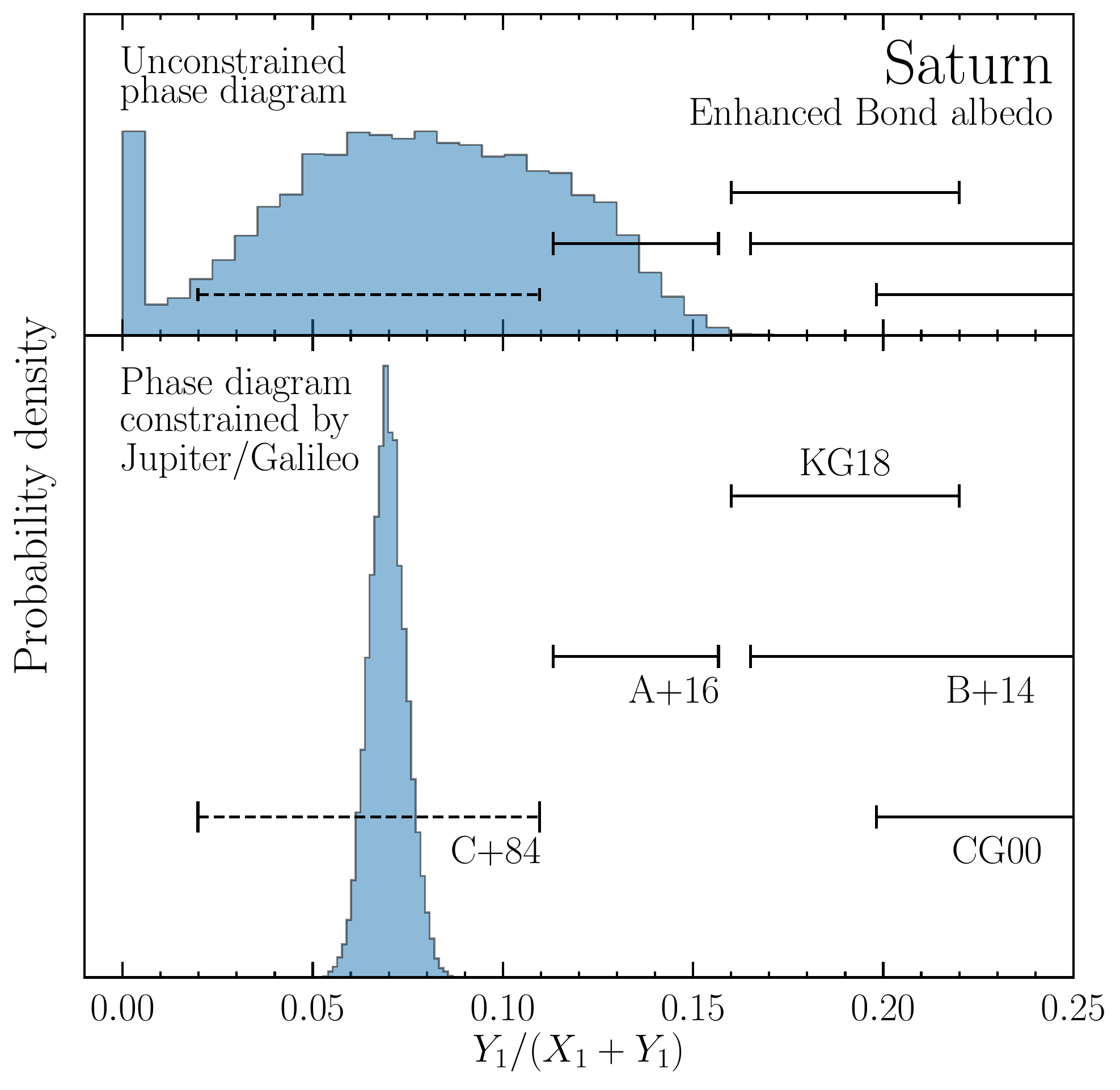} % {../../figures/sat_y1_predictions_emphasize_bottom}
        \caption{
        The helium mass fraction (relative to hydrogen and helium) predicted for Saturn's atmosphere today. \emph{Top:} Saturn models with uniform priors as described in the text. \emph{Bottom:} Saturns with a prior on $\dt$ set by the posterior obtained for Jupiter (Figure~\ref{fig.jup_dt_y1}) and driven by the \galileo probe helium abundance measurement \citep{1998JGR...10322815V}. Black error bars are values derived from \voyager \citep{1984ApJ...282..807C,2000Icar..144..124C} and \cassini \citep{2014DPS....4651201B,2016DPS....4850801A,2018Icar..307..161K} data.
        }
        \label{fig.sat_y1}
    \end{center}
\end{figure}

The abundance of helium in Saturn's atmosphere is the central testable prediction of the Saturn models presented here. Figure~\ref{fig.sat_y1} summarizes our findings in this respect, with our best guess represented by the distribution in the lower panel. Our predicted atmospheric helium mass fraction at $Y_1/(X_1+Y_1)=0.07\pm0.01$ is lower than the earlier theoretical predictions $0.11-0.21$ by \cite{1999P&SS...47.1175H} and $0.13-0.16$ by \cite{2015MNRAS.447.3422N}, a result of the updated phase diagram and hydrogen EOS considered here. The broad distribution in the upper panel are disfavored for the reasons described in Section~\ref{s.sat_dt}.

Observational determinations of the He to H$_2$ mixing ratio have been made by various means, typically combining thermal emission spectra with vertical temperature profiles obtained from radio occultations or infrared limb scans. Values derived in this way from \pioneer \citep{1980JGR....85.5871O}, \voyager \citep{1984ApJ...282..807C}, and \cassini data \citep{2014DPS....4651201B,2016DPS....4850801A,2018Icar..307..161K}, and also from purely infrared \voyager data \citep{2000Icar..144..124C}, have yet to reach a consensus, but they are all consistent with depletion from the protosolar helium abundance. Figure~\ref{fig.sat_y1} compares these values alongside the theoretical results derived here. We predict a more pronounced depletion than implied by all of these observations, with the exception of low estimate of \cite{1984ApJ...282..807C} which may unreliable for reasons explained in \citealt{2000Icar..144..124C}. Given the challenging systematics involved with these measurements, a definitive validation or exclusion of our model may have to await an \emph{in situ} measurement of Saturn's atmospheric helium abundance.

\subsection{Is neon depletion energetically significant?}\label{s.neon}
Besides just helium, Jupiter shows evidence for depletion of its atmospheric neon, exhibiting an abundance around 1/10 the protosolar value by number \citep{1998JGR...10322831N}. This depletion is generally understood to be a consequence of neon's tendency to dissolve into the helium-rich droplets \citep{roulston_stevenson_1995,2010PhRvL.104l1101W} that are lost to the interior. Thus the atmospheric neon depletion observed \emph{in situ} at Jupiter offers a compelling secondary confirmation of the notion that helium differentiation has occurred in Jupiter.

What has not been considered is the energetic significance of sinking neon along with the sinking helium. Assuming that Jupiter's global neon enrichment is similar to its observed atmospheric enrichment in the other noble gases at 2 to 3 times protosolar \citep{2016arXiv160604510A}, then Jupiter's atmospheric neon has depleted by a factor of 20-30. If we further make the assumption that neon was initially well mixed throughout the envelope after formation, and its atmospheric depletion is driven entirely by loss into helium-rich droplets at the molecular-metallic interface, then the degree of neon differentiation at the solar age is simply set by the relative masses in the helium-poor (molecular) and helium-rich (metallic) regions of the interior.

\begin{figure}[t]
    \begin{center}
        \includegraphics[width=1.0\columnwidth]{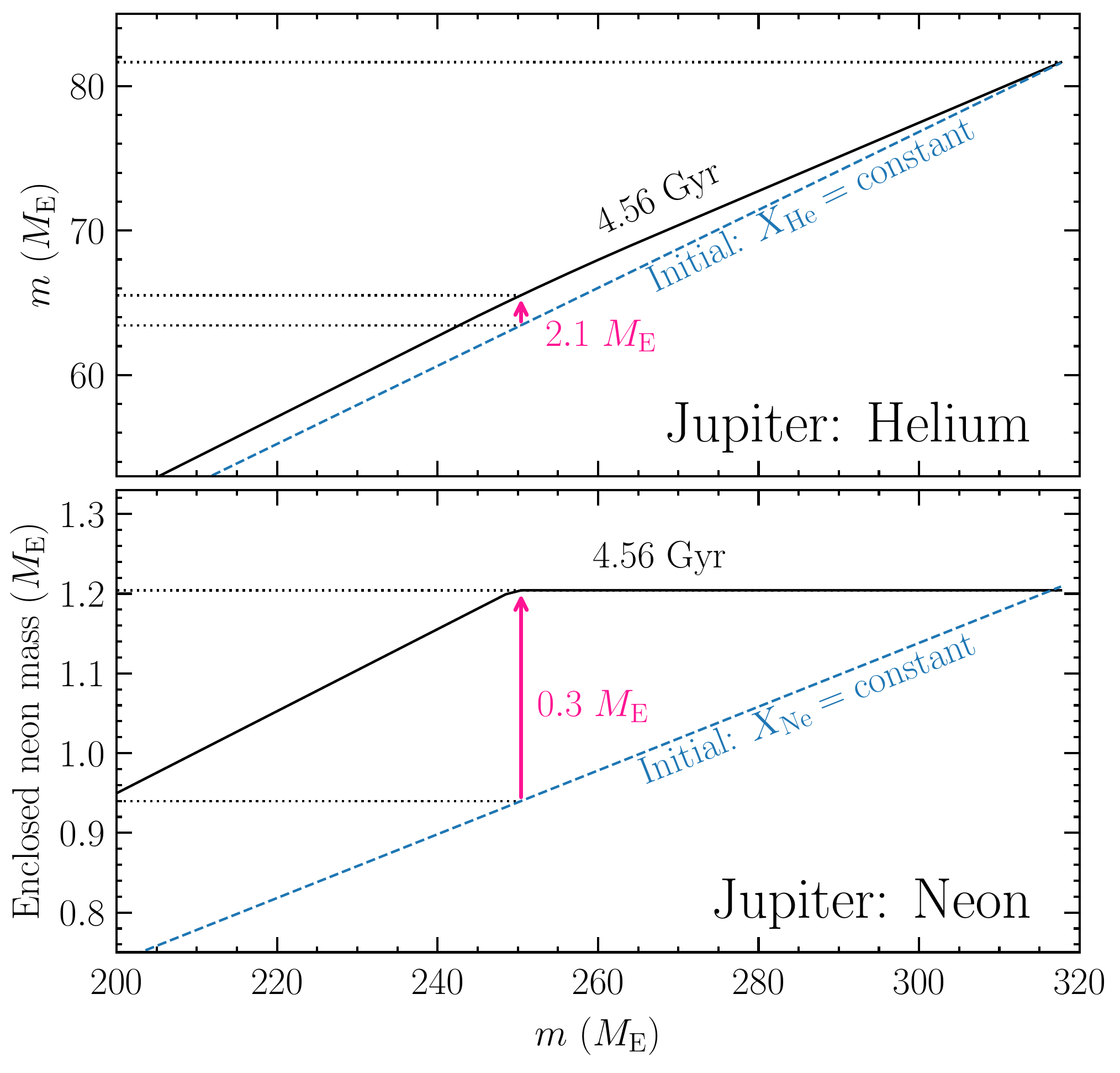} % {../../figures/he_ne_sequestered_jupiter}
        \caption{The initially well-mixed versus final, differentiated, distributions of helium (top panel) and neon (bottom panel) in Jupiter's interior for the simple models discussed in Section~\ref{s.neon}. Enclosed helium (or neon) mass is plotted as a function of mass coordinate. The arrows and adjacent labels indicate the difference between initial and final helium (or neon) mass residing in the molecular envelope; this mass difference is lost to the deeper metallic interior starting with the onset of helium immiscibility.
        }
        \label{fig.jup_neon}
    \end{center}
\end{figure}

Figure~\ref{fig.jup_neon} applies this reasoning to our most likely Jupiter model, showing the enclosed helium or neon mass as a function mass coordinate in the planet. We suppose that Jupiter's bulk neon enrichment is similar to its atmospheric argon enrichment at $\approx3$ times protosolar \citep{2000JGR...10515061M,2009ARA&A..47..481A}, implying an initial neon mass fraction $X_{\rm Ne}\approx4\times10^{-3}$ in Jupiter's envelope for a total neon mass of $M_{\rm Ne}\approx1.2\ M_{\rm E}$, approximately $0.3\ M_{\rm E}$ of which resides in the molecular envelope when helium immiscibility sets in. The observed atmospheric abundance at about 1/10 protosolar---the exact value is taken from \citealt{2000JGR...10515061M}---translating into a final neon mass fraction of $X_{\rm Ne}\approx2\times10^{-4}$ in Jupiter's molecular envelope for a final neon mass $M_{\rm Ne}\approx10^{-2}\ M_{\rm E}$ remaining there. The molecular envelope thus lost virtually all $0.3\ M_{\rm E}$ of its neon since the onset of helium immiscibility, compared to the $\approx2.1\ M_{\rm E}$ of helium that sank to the metallic depths for the same model.

\begin{figure}[t]
    \begin{center}
        \includegraphics[width=1.0\columnwidth]{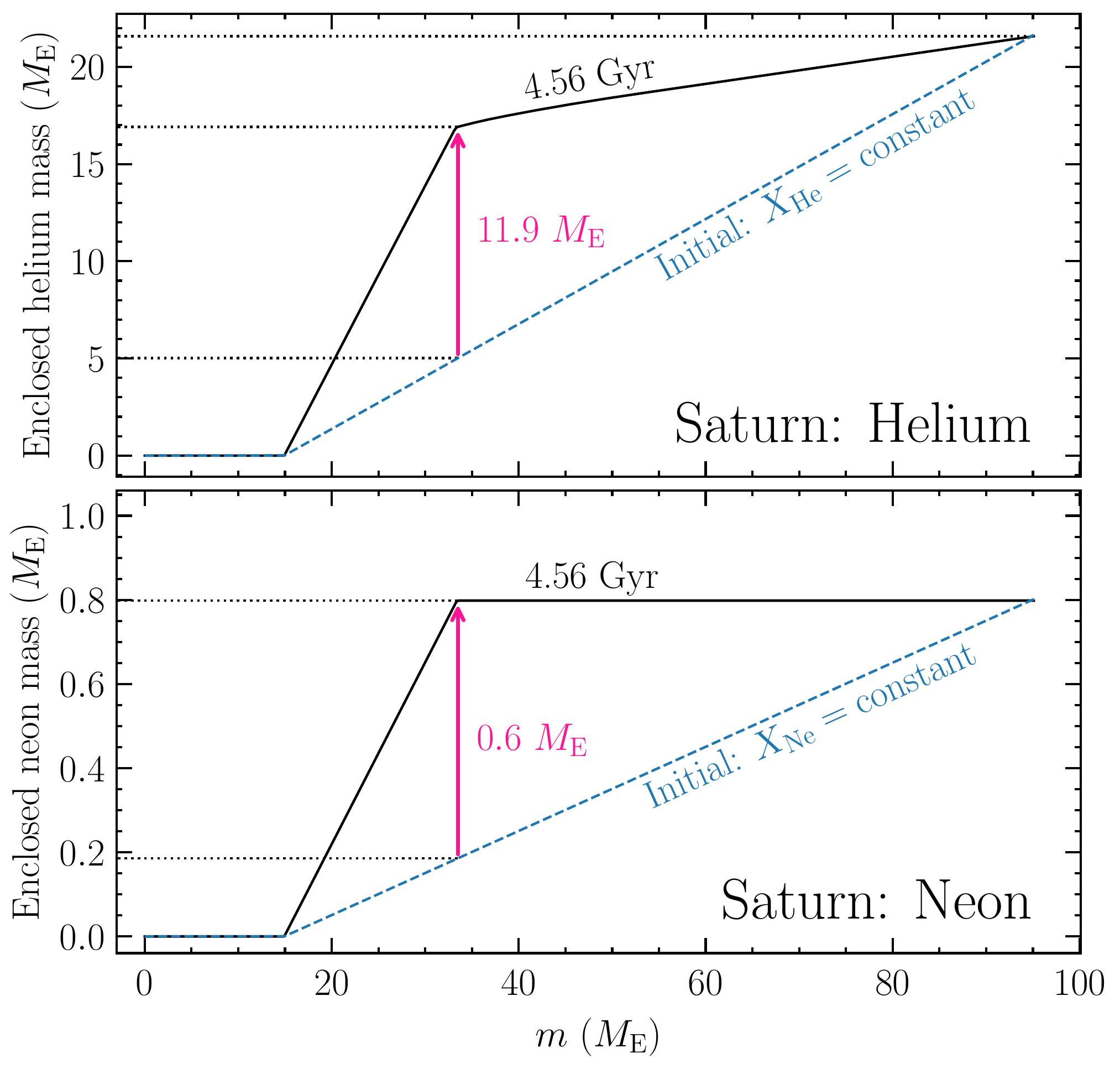} % {../../figures/he_ne_sequestered_saturn}
        \caption{As in Figure~\ref{fig.jup_neon}, but for Saturn.
        }
        \label{fig.sat_neon}
    \end{center}
\end{figure}

Our models suggest that the differentiation of helium is more advanced in Saturn, and consequently depletion of neon in Saturn's atmosphere may be even more advanced than in Jupiter's. For the sake of these simple estimates, we assume that Saturn's outer envelope contains a negligible mass of neon at the present day. As before we assign a fiducial bulk enrichment for neon based on measurements of different species, this time supposing that neon tracks the carbon enrichment at $\sim10\times$ protosolar per the methane abundance from \cite{2009Icar..199..351F}. We assume that the dissolved neon follows helium-rich droplets all the way to the helium-rich shell; the neon transition in this simplistic model therefore takes place substantially deeper than the molecular-metallic transition. Figure~\ref{fig.sat_neon} illustrates the result of applying this exercise to our most likely Saturn model including the Jupiter/\galileo phase diagram prior and enhanced Bond albedo. This model sheds $0.6\ M_{\rm E}$ of neon from its outer envelope compared to $11.9\ M_{\rm E}$ of helium.

Calculating the associated change in the gravitational binding energy of neon provides an estimate of the energetic significance of neon differentiation. From the composition profiles in Figures~\ref{fig.jup_neon} and \ref{fig.sat_neon} we calculate initial and final values for the binding energies
\begin{equation}
  E_{\rm He}=-\int_0^{M_{\rm He}}\frac{Gm}{r}\,{\rm d}m_{\rm He}
\end{equation}
and
\begin{equation}
  E_{\rm Ne}=-\int_0^{M_{\rm Ne}}\frac{Gm}{r}\,{\rm d}m_{\rm Ne}
\end{equation}
with $m_{\rm He}$ the enclosed helium mass, $m_{\rm Ne}$ the enclosed neon mass, and $m$ the enclosed total mass. For Jupiter, we find that $\Delta E_{\rm Ne}\sim10^{40}\ {\rm erg}$ compared to $\Delta E_{\rm He}\sim10^{41}\ {\rm erg}$. In the case of Saturn, we find $\Delta E_{\rm Ne}\sim3\times10^{39}\ {\rm erg}$ compared to $\Delta E_{\rm He}\sim10^{41}\ {\rm erg}$. We thus expect that neon sequestration could bolster the luminosity from helium differentiation by as much as $\sim10\%$ for Jupiter and $\sim3\%$ for Saturn, an effect not captured in the thermal evolution models of this work.

These crude arguments suggest that the luminosity that helium differentiation provides to each planet is augmented somewhat by the accompanying sequestering of neon, at least in a time-averaged sense. Apart from the unknown bulk neon abundances for Jupiter and Saturn, there is an additional layer of complexity associated with the precise time evolution. Considering Jupiter's pronounced neon depletion in spite of helium immiscibility having set in only recently, it seems possible that Saturn's neon was sequestered rapidly after immiscibility set in $\gtrsim3$ Gyr ago. In this case the luminosity source may have been episodic in nature---potentially on timescales close to the thermal timescale---and deducing its influence on Saturn's cooling time as a whole would require a more detailed thermal evolution model as well as a more detailed hydrogen-helium-neon immiscibility model. For these reasons, explicit treatment of neon in the evolutionary models is beyond the scope of this work, but deserves closer attention as models like these are refined.

\section{Discussion}\label{s.discussion}
\subsection{Implications}
This work provides a self-consistent physical picture for the thermal evolution of Jupiter and Saturn in detail, an outcome that had so far proven elusive. These models are built on the premise that some degree of hydrogen-helium immiscibility---and rainout of the resulting helium-rich phase---occurs in both planets, a notion supported by decades of work spanning dense matter physics \citep[e.g.,][]{1975PhRvB..12.3999S,1985ApJ...290..388H,2009PNAS..106.1324M,2009PhRvL.102k5701L} and planetary science \citep[e.g.,][]{1967Natur.215..691S,ss77b,1998JGR...10322815V,2000Icar..144..124C}. The evolution models presented here apply recent advances in the equation of state of hydrogen \citep{2013ApJ...774..148M}, the phase diagram describing miscibility of hydrogen-helium mixtures \citep{2018PhRvL.120k5703S}, and the atmospheres of the gas giants \citep{fortney2011,2018NatCo...9.3709L}. Sampling parameter space systematically using Markov chain Monte Carlo, we are able to arrive at solutions that naturally explain the radii and heat flow of both Jupiter and Saturn at the solar age, as well as Jupiter's observed atmospheric helium depletion.

The value in this final constraint is that it puts stringent limits on allowable phase diagrams. The parameter estimation performed in this work provides statistically meaningful distributions of model parameters, estimating for instance that based on Jupiter's helium depletion, the true phase diagram is warmer than the most current \emph{ab initio} phase diagram \citep{2018PhRvL.120k5703S} by $(539\pm23)\ {\rm K}$ ($1\sigma$ uncertainty) at the $\approx2\ {\rm Mbar}$ pressures that it predicts for the onset of helium immiscibility in metallic hydrogen. For comparison, Jupiter models built on a previous generation of phase diagram that assumed ideal hydrogen-helium mixing entropy predicted the necessary temperature offset to be between 200 and 300 K, in the opposite direction \citep{2015MNRAS.447.3422N,2016ApJ...832..113M}. These findings imply a rather precise prediction for Saturn's atmospheric helium abundance summarized in Figure~\ref{fig.sat_y1}. The posterior predictive $Y_1$ distribution for our favored model is well fit by a Gaussian producing a helium mass fraction, relative to hydrogen and helium, of $0.07\pm0.01$ ($2\sigma$ uncertainty). The corresponding $\rm{He/H}_2$ mixing ratio is $0.036\pm0.006$, consistent with one measurement made from multi-instrument \cassini data but inconsistent with others. An \emph{in situ} determination of Saturn's atmospheric helium abundance provided by an entry probe \citep[e.g., SPRITE;][]{2018AAS...23114401S} would be decisive test of the evolutionary picture developed here.

Although superadiabaticity resulting from some flavor of double-diffusive convection in metallic regions possessing helium gradients does modulate the cooling time for Jupiter and Saturn, it is not required in all cases. In particular, good solutions for Jupiter do require $\rrho>0$, consistent with the expectation from earlier modeling efforts \citep[e.g.,][]{1999P&SS...47.1175H,2003Icar..164..228F,fortney2011,2015MNRAS.447.3422N,2016ApJ...832..113M} that Jupiter required some mechanism (non-adiabatic interiors or otherwise) to speed its evolution rather than prolong it\footnote{Recognizing this tension, \cite{2016ApJ...832..113M} ultimately treated Jupiter's Bond albedo as a free parameter, recovering a median value consistent with the subsequent \cassini measurement.}. However, the new models find interiors that are closer to adiabatic ($\rrho$ closer to zero) due to the major improvement in our understanding of the internal heat flow of Jupiter in light of results from \cassini \citep{2018NatCo...9.3709L}. For Saturn, equally good solutions are found assuming purely adiabatic envelopes corresponding to essentially perfect convection, although nonzero values $\rrho\approx0.05$ typical of our Jupiter models (Figure~\ref{fig.jup_rrho_age}) are also likely in our Saturn models (Figure~\ref{fig.sat_rrho_age}).

Finally the energetic significance of neon differentiation is examined. Assuming that Jupiter's atmospheric neon depletion is driven by dissolution into the helium-rich material lost to the metallic interior, and making an informed guess about the planet's bulk neon abundance, we estimate that Jupiter's time-averaged differentiation luminosity may be increased by $\sim10\%$ relative to just the helium differentiation treated in the thermal evolution models in this work. Assuming similar bulk abundance patterns for Saturn and assuming that its outer envelope is devoid of neon by the solar age, Saturn's differentiation luminosity could be augmented by $\sim3\%$.

\subsection{Constraints not addressed in this work}
The one-dimensional evolutionary models constructed here neglect rotation and so do not incorporate constraints from Jupiter or Saturn's observed gravity fields and shape, invaluable constraints on the interiors of both planets. The Jupiter models presented above are quite similar to those already in the literature \citep[e.g.,][]{2018Natur.555..227G}. The helium shells present in our Saturn models, on the other hand, yield quite different mass distributions than are usually considered in the course of intepreting Saturn's gravity field \citep[e.g.,][]{2019ApJ...879...78M,2019Sci...364.2965I}. The implications of this alternative structure for Saturn as it pertains to the \cassini gravity field are addressed in a companion article that focuses on static models consistent with the end state of the Saturn evolutionary models presented above.

The models in this work allow for statically stable regions associated with stabilizing helium gradients provided by helium rainout; physically these regions correspond to superadiabatic, double-diffusive regions. However, the models parsimoniously assume simple metallicity distributions---distinct cores and envelopes homogeneous in $Z$---even though these configurations may not be well justified. For example, a dilute core has been argued to exist in Jupiter on the basis of the \juno gravity measurements \citep{2017GeoRL..44.4649W} and models of formation \citep{2017ApJ...840L...4H} or an early giant impact \citep{2019Natur.572..355L}; {\cite{2019ApJ...872..100D} also propose an innovative model for Jupiter's interior appealing to double-diffusive convection.}  \emph{At least} one statically stable region has been argued to exist in Saturn's interior from the independent perspectives of Kronoseismology \citep{2014Icar..242..283F} and dynamo models \citep[e.g.,][]{1982GApFD..21..113S,2011E&PSL.304...22C,2019arXiv191106952C}; it is unclear whether these stable stratifications are supported by helium gradients, metallicity gradients, or some combination of the two. It is beyond our present scope to explore the additional complexities of more complicated metallicity distributions, and certainly to fully analyze the implications that these helium distributions have for the planets' free oscillations and dynamo generation. Thus, while this work realizes a compelling pathway for the evolution of Jupiter and Saturn together, the present model is necessarily incomplete.

\section{Conclusions}\label{s.conclusions}
An explanation for Saturn's surprisingly high luminosity has been sought for decades. Models that provide plausible evolution pathways for Saturn invoke either an additional luminosity source beyond straightforward cooling, or interiors that deviate significantly from adiabaticity because of some degree of non-convective heat transport. On the other hand, Jupiter's luminosity is fairly well explained by simple models, but its empirically well-constrained atmospheric abundances reveal the presence of interior processes that sequester helium and neon. This work applies identical assumptions to Jupiter and Saturn, calculating new thermal evolution models in the context of recent results regarding hydrogen-helium immiscibility physics and a significantly revised measurement of Jupiter's intrinsic heat flow. We have shown that these models naturally address the observed heat flow from both Jupiter and Saturn at the solar age, as well as Jupiter's atmospheric helium depletion. Our main findings are that
\begin{enumerate}
  \item {Jupiter's observed atmospheric helium depletion \citep{1998JGR...10322815V} implies that hydrogen-helium phase separation sets in $(539\pm23\ {\rm K})$ warmer than the phase curves of \cite{2018PhRvL.120k5703S} predict at $P=2\ {\rm Mbar}$ (Figure~\ref{fig.jup_dt_y1});}
  \item {Jupiter's heat flow through its thin helium gradient region is somewhat superadiabatic at $\rrho\approx0.05$ (Equation~\ref{eq.rrho}; Figure~\ref{fig.jup_rrho_age});}
  \item {Phase diagrams covering helium mixing ratio space are necessary to self-consistently predict the helium distribution within a cool gas giant, as made clear here for Saturn where differentiation leads to large local departures from a protosolar or Jupiter-like mixture (Section~\ref{s.mixtures});}
  \item {In the limit of rapid rainout of helium overdensities considered here, realistic phase diagrams invariably predict a dense, helium-rich layer deep within Saturn (Figures~\ref{fig.sat_phase} and \ref{fig.pie_slices});}
  \item {Saturn models satisfying phase diagrams consistent with the Jupiter/\galileo helium constraint fail to reproduce Saturn's observed heat flow at the solar age if the \voyager-era estimate for Saturn's Bond albedo $A\sim0.3$ holds (Figure~\ref{fig.sat_dt_age});}
  \item {Saturn models that are successful in all respects are found if Saturn's true Bond albedo is $A\sim0.5$ (Figures~\ref{fig.sat_dt_high_albedo_tracks}-\ref{fig.sat_dt_y1_high_albedo});}
  \item {Adiabatic and superadiabatic interiors are equally likely for Saturn (Figure~\ref{fig.sat_rrho_age});}
  \item {Neon differentiation probably makes a significant energetic contribution to these planets' thermal evolution, albeit one to two orders of magnitude weaker than helium differentiation (Section~\ref{s.neon});}
  \item {The phase diagrams consistent with the Jupiter/\galileo helium constraint make a precise prediction for Saturn's atmospheric helium abundance, predicting a strong depletion at $Y=0.07\pm0.01$ for a He/H$_2$ mixing ratio $0.036\pm0.006$ (Figure~\ref{fig.sat_y1}).}
\end{enumerate}
Measuring Saturn's atmospheric helium abundance \emph{in situ} would provide an observational test of the picture developed here.

\acknowledgements
The authors give thanks to the anonymous referee for helpful critiques, and to M. Sch\"ottler for providing the phase diagram data used for this work.
This work was supported by NASA through Earth and Space Science Fellowship program grant NNX15AQ62H to C.M. and Cassini Participating Scientist program grant NNX16AI43G to J.J.F.

\software{
  \texttt{emcee} \citep{2013PSP..125..306F},
  SciPy \citep{scipy},
  NumPy \citep{numpy},
  Matplotlib \citep{Hunter:2007}
}
\facilities{ADS}

% \appendix

\bibliography{bib}
\end{document}